# *Observational Constraints on the Acceleration Discrepancy Problem*

*Stacy McGaugh*

*University of Maryland*

*What gets us into trouble is not what we don't know.*

*It's what we know for sure that just aint so.*

— Mark Twain

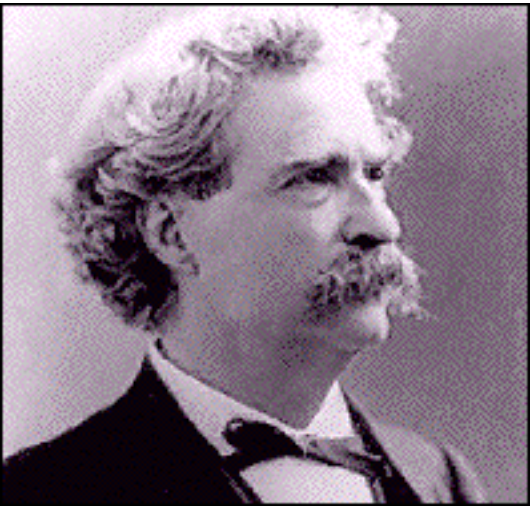

i.e., we shouldn't be overly confident that the universe is filled with some new form of invisible, non-baryonic mass (such as WIMPs) until we actually detect the stuff directly in the laboratory. Current cosmology ($\Lambda$CDM) invokes not one but two aethers (dark matter and dark energy); let us be careful not to fall into the same conceptual trap that led classical physicists to infer that Maxwell's theory *required* aether. It is at least conceivable that there could be a theory which captures the successes of cosmology without the excess baggage.

The Dark Matter Tree:

The roots represent the empirical roots of the problem; the branches the various proposed solutions.

Dark matter solutions are represented on the left branch; modifications of dynamical laws are represented on the right branch.

This review focusses on the phenomenology of the mass discrepancy, which appears at a particular acceleration scale. By request, I focus on rotation curves, but will also touch on other data.
Time restrictions limit discussion of theories to the specific case of MOND.

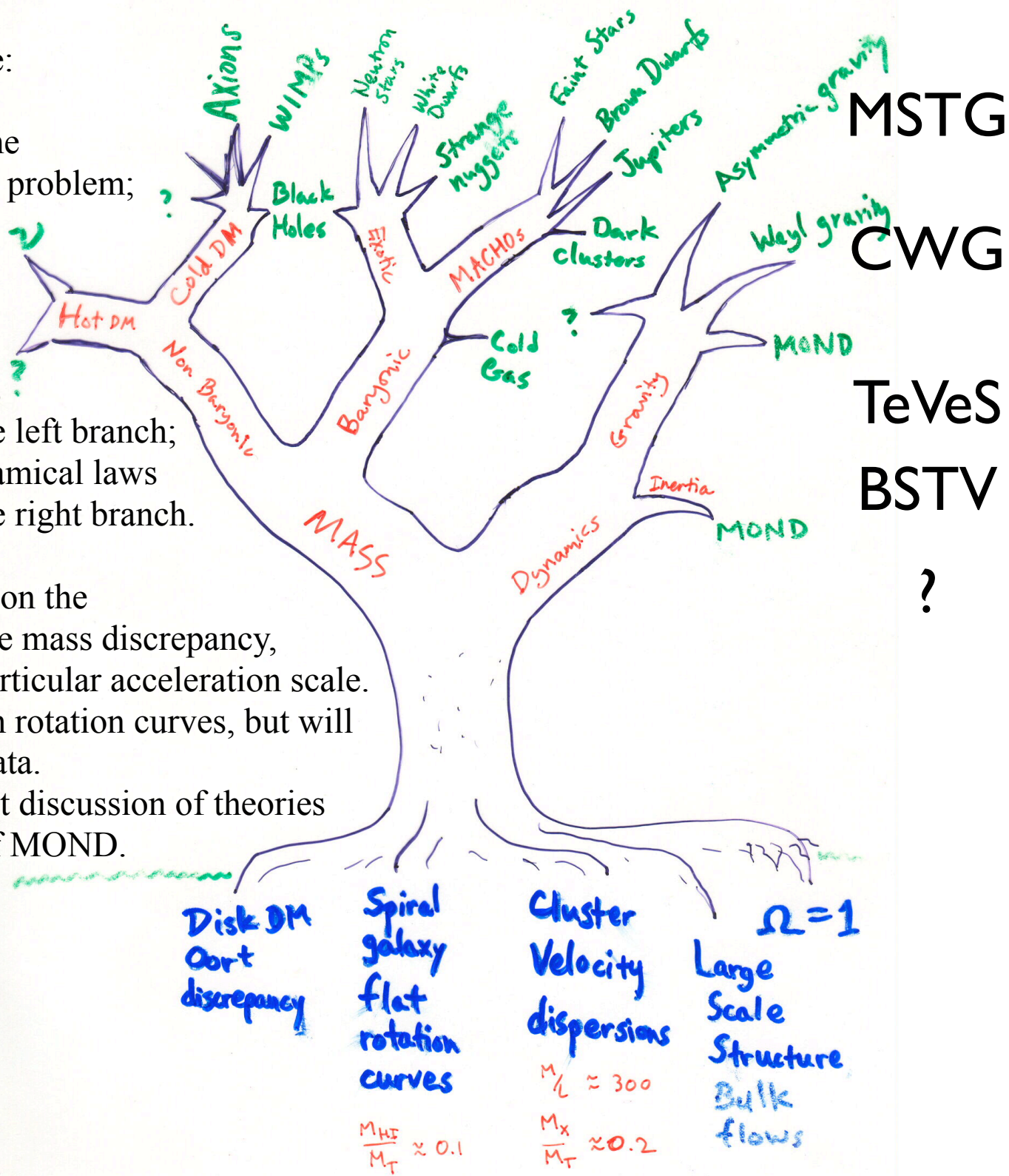

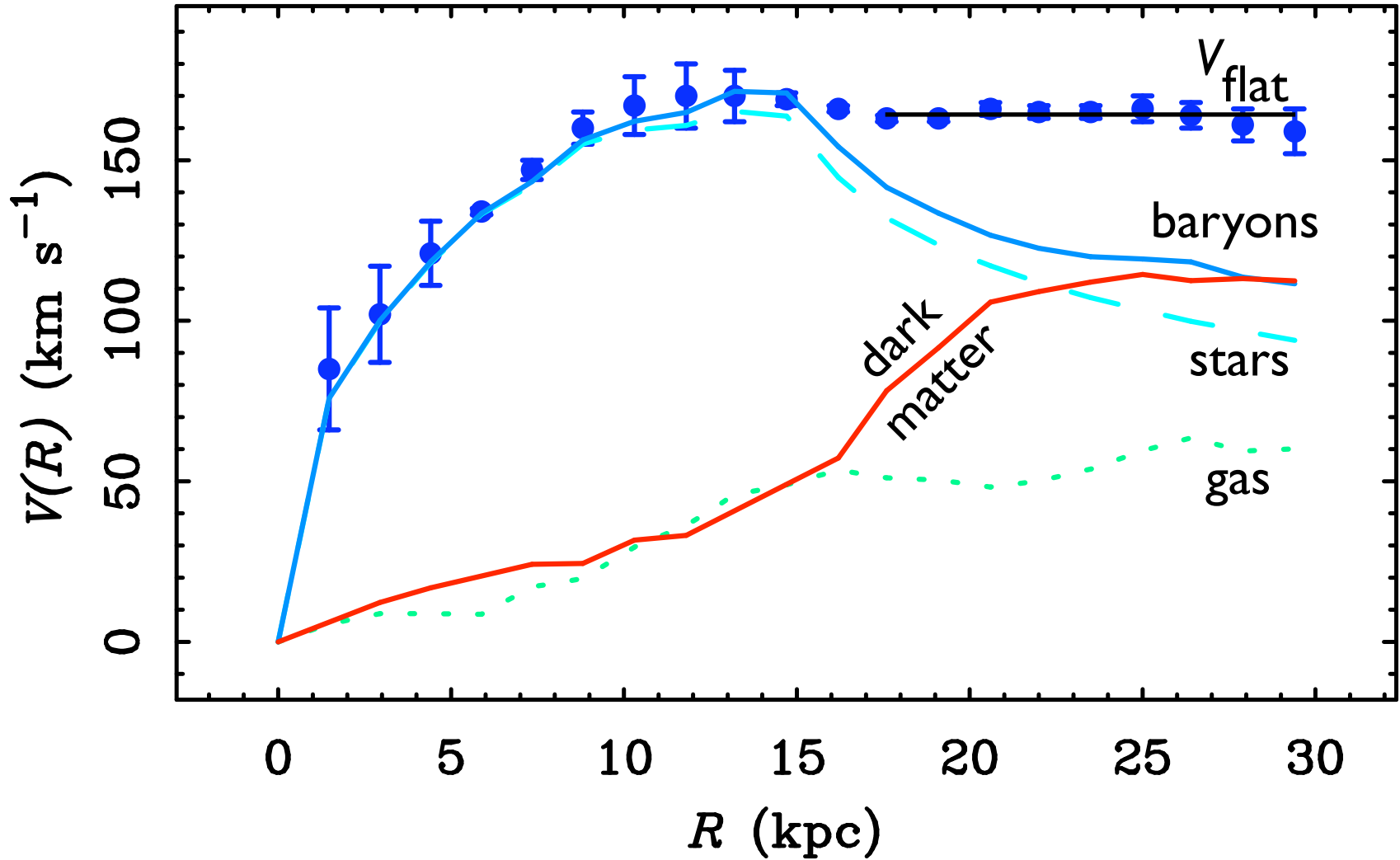

The rotation curve and mass model of NGC 6946 (pictured on the first slide). The baryons are the sum of stars and gas computed from the observed mass distribution for the case of maximum disk. Dark matter is whatever is left over.

# Tully-Fisher

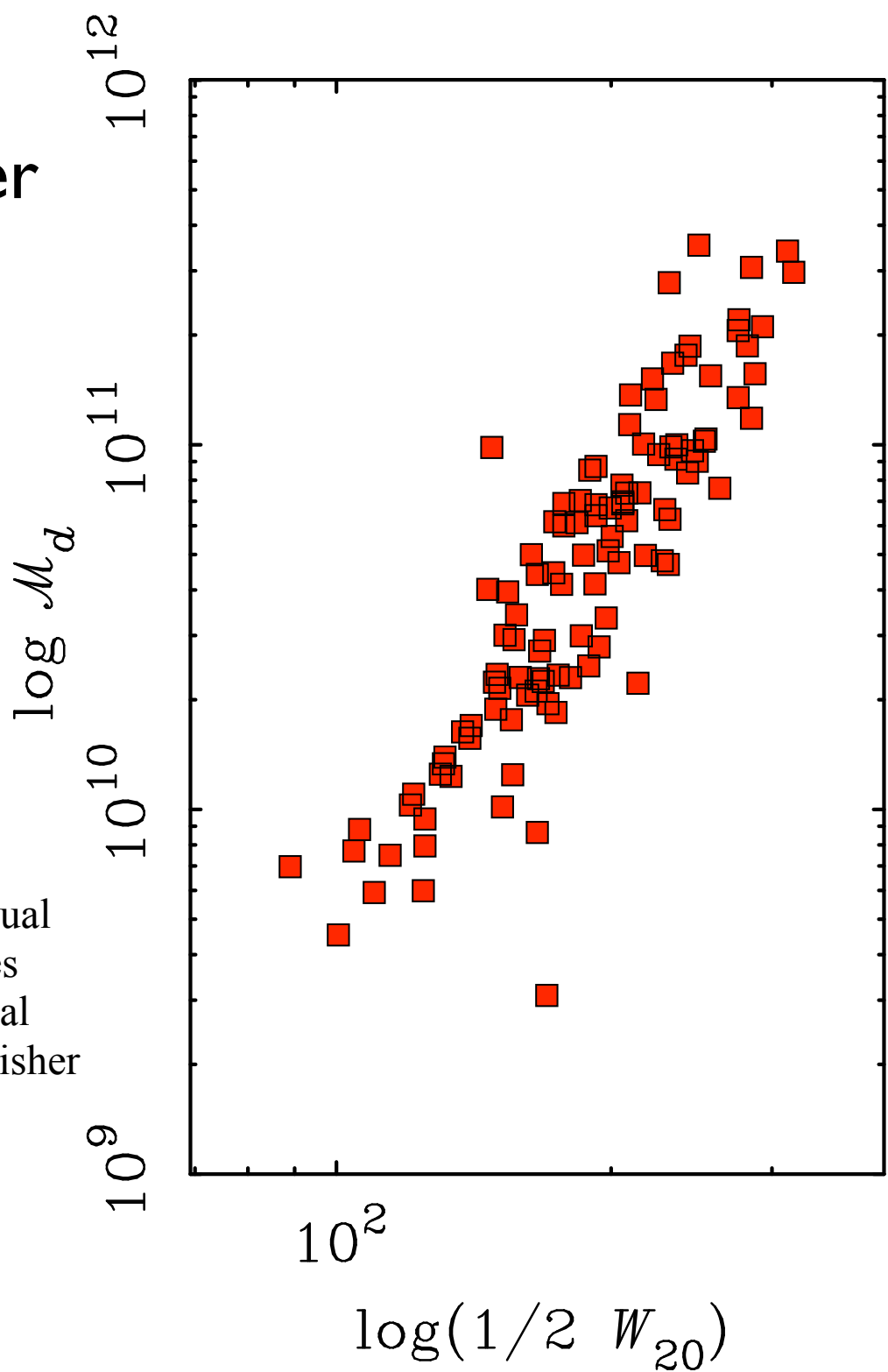

relation between luminosity/mass and rotation speed

Bothun et al. (1985)
*H*-band

In addition to individual mass models, galaxies adhere to strong global relations like Tully-Fisher

## TF Relation

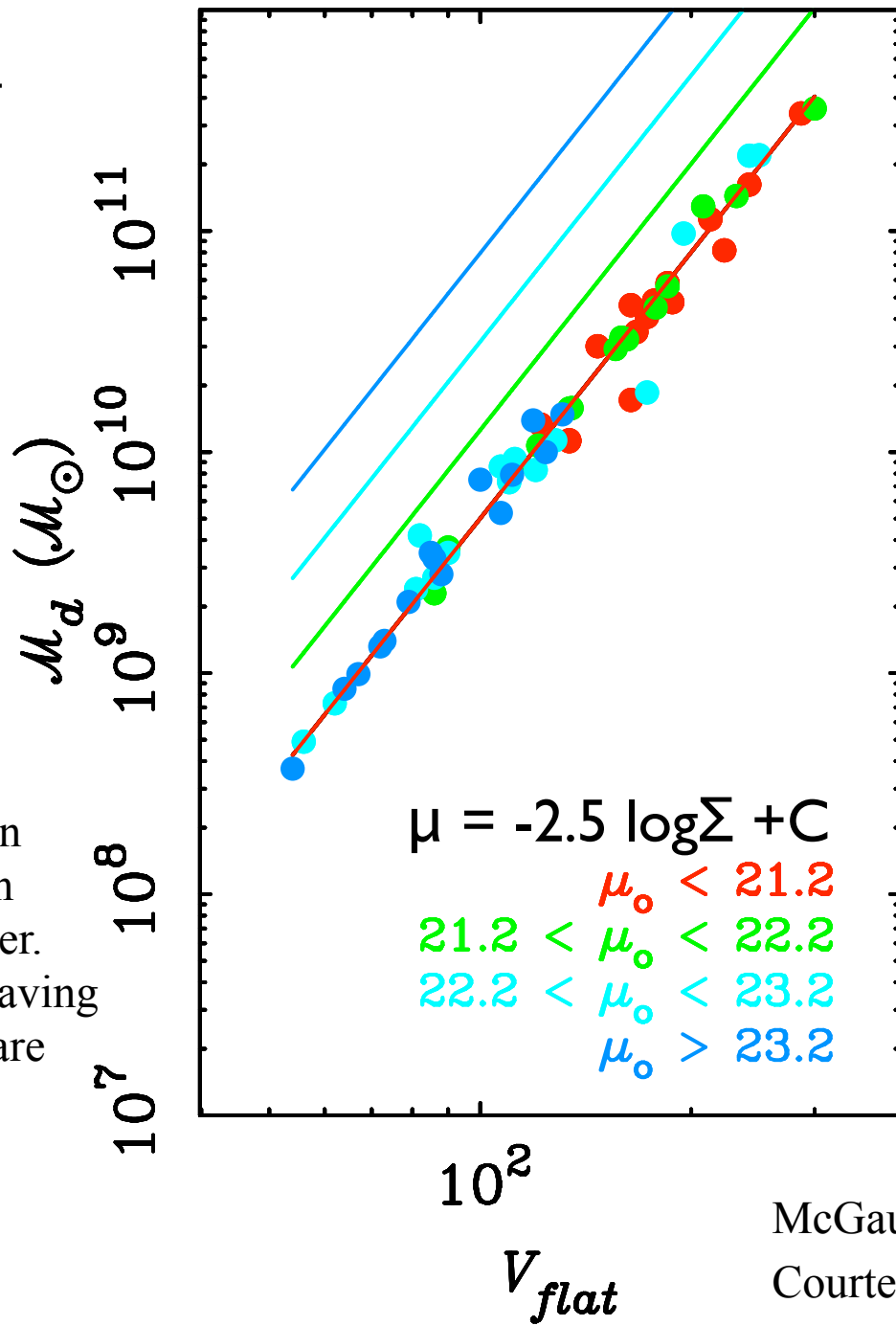

**Newton says**
$V^2 = GM/R$.
Equivalently,
$\Sigma = M/R^2$
$V^4 = G^2 M \Sigma$

**Therefore**
Different $\Sigma$ *should* mean different TF normalization.

$\mu = -2.5 \log \Sigma + C$
$\mu_o < 21.2$
$21.2 < \mu_o < 22.2$
$22.2 < \mu_o < 23.2$
$\mu_o > 23.2$

The Tully-Fisher relation is not well understood in the context of dark matter. There are many hand-waving models, none of which are completely satisfactory.

One expects, from basic physics, that variations in the distribution of baryonic mass should have an impact on the Tully-Fisher relation (lines). They do not (data). See discussions in

McGaugh & de Blok 1998, ApJ, 499, 41
Courteau & Rix 1999, ApJ, 513, 561

# No Residuals from TF rel'n

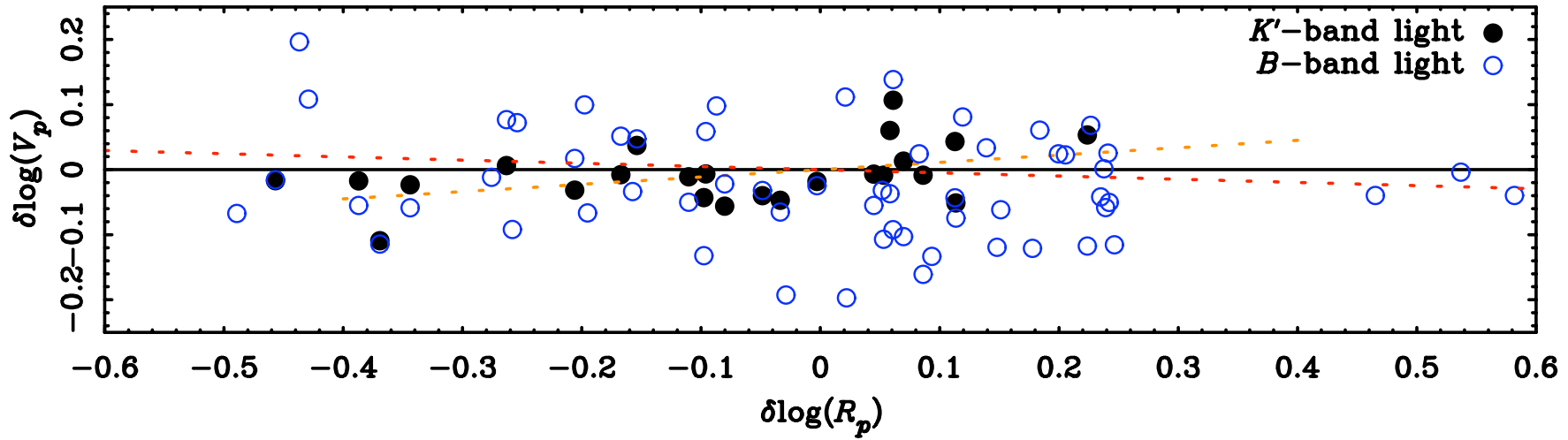

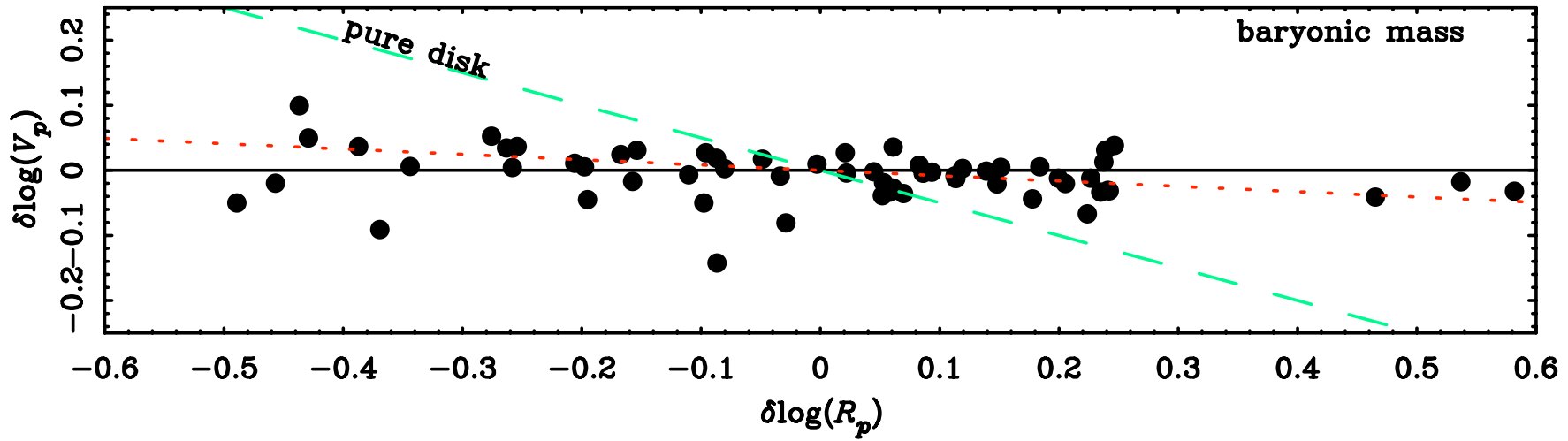

Indeed, the residuals from Tully-Fisher are nearly to totally imperceptible, depending weakly on the choice of circular velocity measured.
This causes a fine-tuning problem which is generic to any flavor of dark matter...

# Requires fine balance between dark & baryonic mass

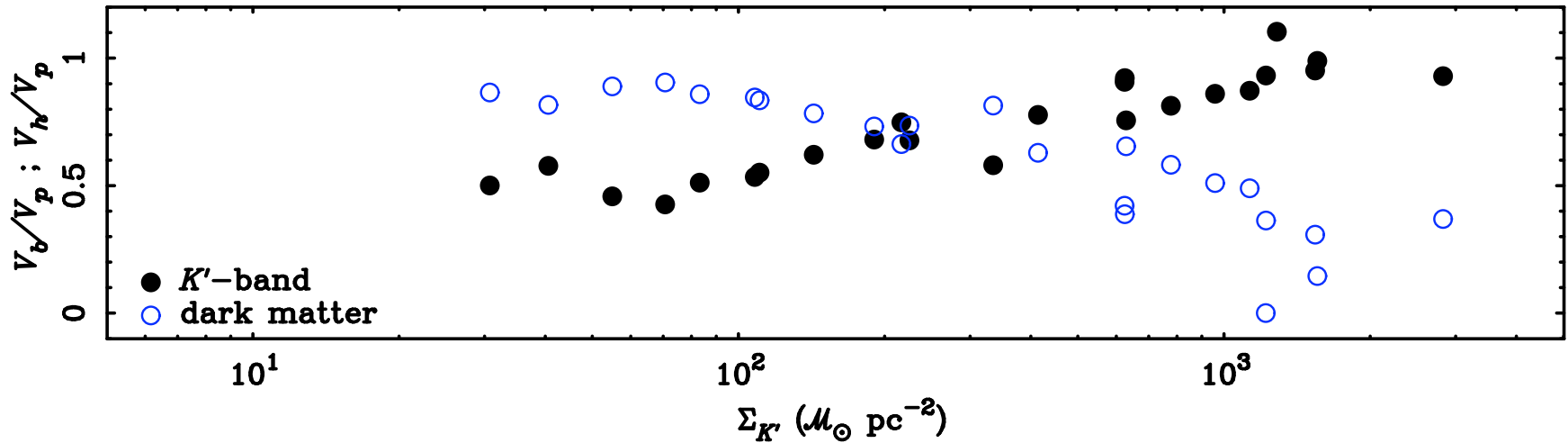

McGaugh 2005, Phys. Rev. Lett. 95, 171302

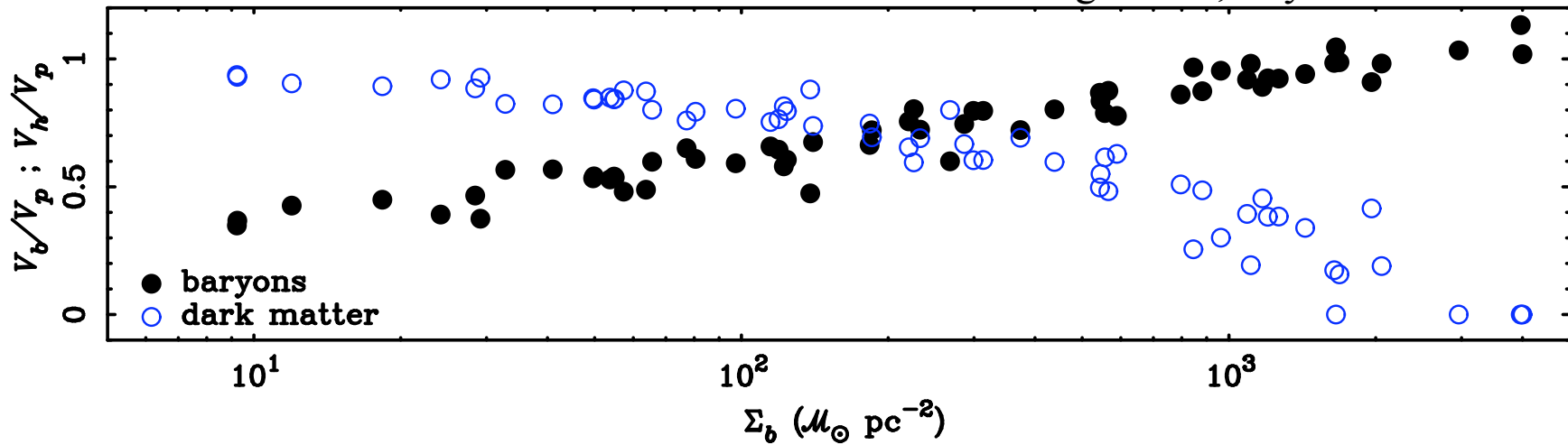

The contribution of the baryonic (filled points) and dark matter (open points) to any given point along the rotation curve must be finely balanced, like a see-saw. As the baryonic contribution increases with baryonic surface density, the dark matter contribution decreases. The two components know intimately about each other...

# Renzo's Rule:

*"When you see a feature in the light, you see a corresponding feature in the rotation curve."*

(Sancisi 1995, private communication)

(See also Sancisi 2004, IAU 220, 233)

*The distribution of mass is coupled to the distribution of light.*

## Quantify by defining the Mass Discrepancy:

$$\mathcal{D} = \frac{V^2}{V_b^2} = \frac{V^2}{\Upsilon_\star v_\star^2 + V_g^2}$$

This is essentially the ratio of total to baryonic mass.
It depends on the stellar mass-to-light ratio, $\Upsilon_\star$,
for which we can explore many possibilities.

The mass discrepancy does not care about a particular length scale. It does correlate strongly with the centripetal acceleration.

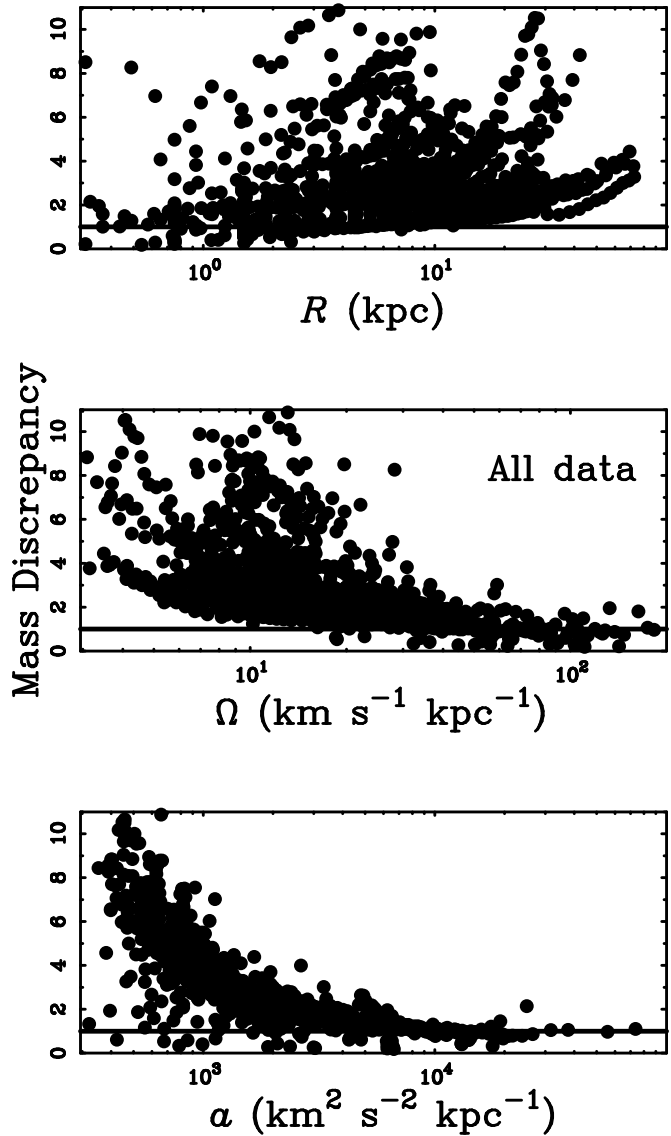
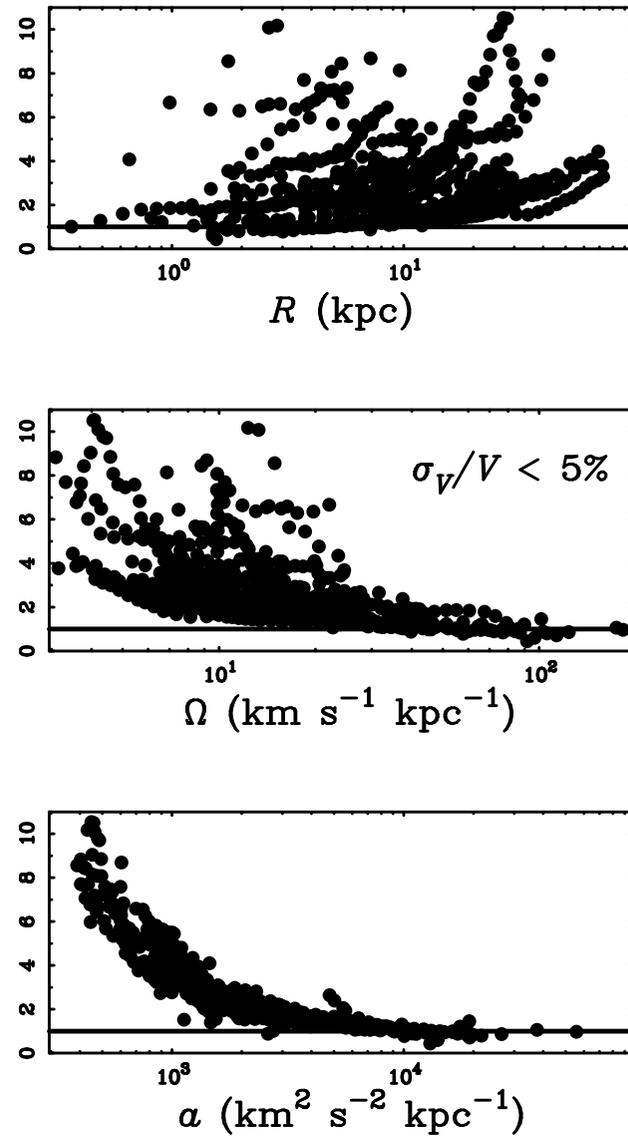

McGaugh 2004, ApJ, 609, 652

radius

orbital frequency

acceleration

74 galaxies
> 1000 points
(all data)

60 galaxies
> 600 points
(errors < 5%)

Before we explore the choice of stellar mass-to-light ratio, note that the basic phenomenology is already present in the data, without any free parameters.

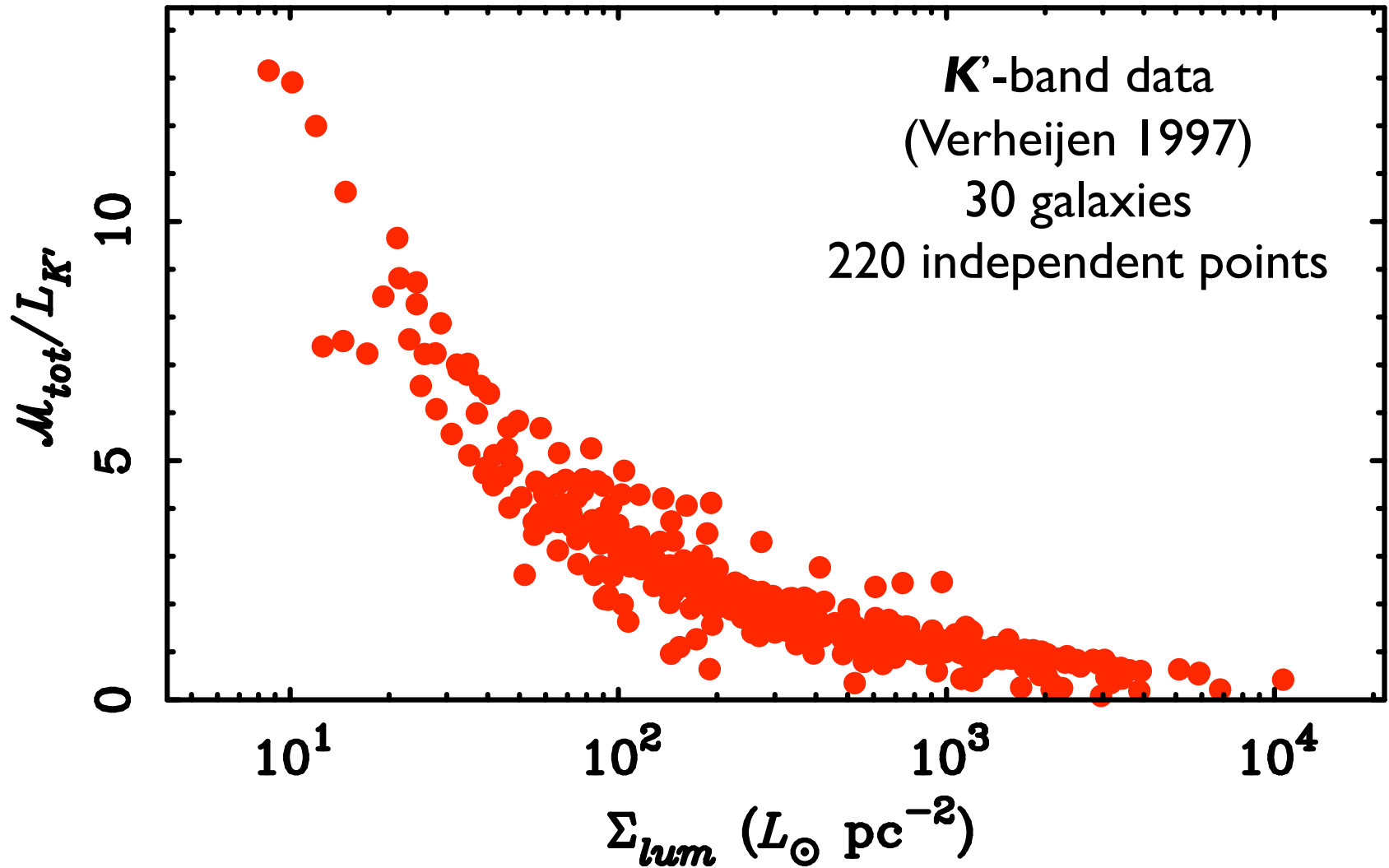

The K'-band dynamical mass-to-light ratio correlates with surface brightness. Note that acceleration relates to surface density: $g_N = G\Sigma$

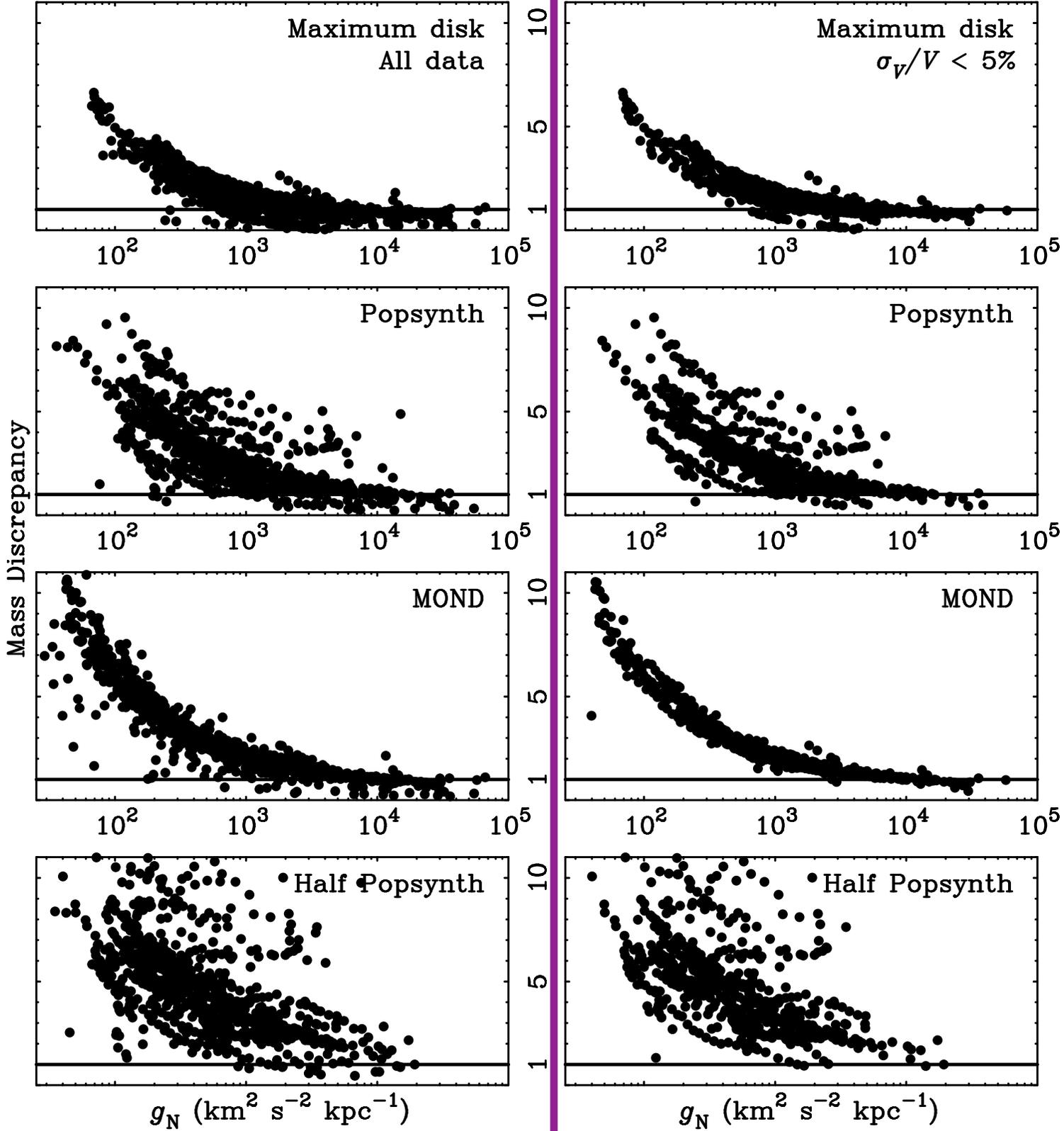

# MOND
## MOdified Newtonian Dynamics
introduced by Moti Milgrom in 1983

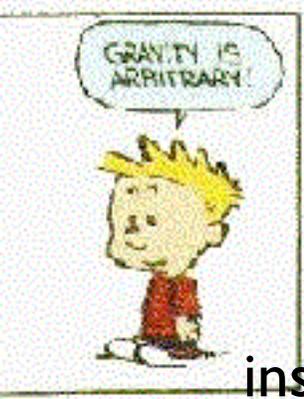

instead of dark matter, suppose the force law changes such that

$$\text{for } a \gg a_o, \ a \Rightarrow g_N$$

$$\text{for } a \ll a_o, \ a \Rightarrow \sqrt{(g_N a_o)}$$

where

$$g_N = GM/R^2$$

is the usual Newtonain acceleration.
More generally, these limits are connected by a smooth interpolation fcn $\mu(a/a_o)$ so that

$$\mu(a/a_o) \ a = g_N.$$

MOND can be interpreted as a modification of either
inertia (F = ma) or gravity (the Poisson eqn).

ApJ, 270, 381

Milgrom 1983

No. 2, 1983  MODIFICATION OF NEWTONIAN DYNAMICS  381

A major step in understanding ellipticals can be made if we can identify them, at least approximately, with idealized structures such as the FRCL spheres discussed above. I have also studied isotropic and nonisotropic isothermal spheres, in the modified dynamics, as such possible structures. I found that they have properties which very much resemble those of ellipticals and galactic bulges. I describe these in Milgrom (1983c).

VIII. PREDICTIONS

The main predictions concerning galaxies are as follows.
1. Velocity curves calculated with the modified dynamics on the basis of the observed mass in galaxies should agree with the observed curves. Elliptical and S0 galaxies may be the best for this purpose since (a) practically no uncertainty due to obscuration is involved and (b) there is not much uncertainty due to the possible presence of molecular hydrogen.
2. The relation between the asymptotic velocity ($V_\infty$) and the mass of the galaxy ($M$) ($V_\infty^4 = MGa_0$) is an absolute one.
3. Analysis of the $z$-dynamics in disk galaxies using the modified dynamics should yield surface densities which agree with the observed ones. Accordingly, the same analysis using the conventional dynamics should yield a discrepancy which increases with radius in a predictable manner.
4. Effects of the modified dynamics are predicted to be particularly strong in dwarf elliptical galaxies (for review of properties see, e.g., Hodge 1971 and Zinn 1980). For example, those dwarfs believed to be bound to our Galaxy would have internal accelerations typically of order $a_{in} \sim a_0/30$. Their (modified) acceleration, $g$, in the field of the Galaxy is larger than the internal ones but still much smaller than $a_0$, $g \approx (8 \text{ kpc}/d)a_0$, based on a value of $V_\infty = 220$ km s$^{-1}$ for the Galaxy, and where $d$ is the distance from the dwarf galaxy to the center of the Milky Way ($d \sim 70$–220 kpc). Whichever way the external acceleration turns out to affect the internal dynamics (see the discussion at the end of § II, the section on small groups in Paper III, and Paper I), we predict that when velocity dispersion data is available for the dwarfs, a large mass discrepancy will result when the conventional dynamics is used to determine the masses. The dynamically determined mass is predicted to be larger by a factor of order 10 or more than that which can be accounted for by stars. In case the internal dynamics is determined by the external acceleration, we predict this factor to increase with $d$ and be of order ($d/8$ kpc) (as long as $a_{in} \ll g$, $h_{50} = 1$).
Prediction 1 is a very general one. It is worthwhile listing some of its consequences as separate predictions, numbered 5–7 below (note that, in fact, even prediction 2 is already contained in prediction 1).

5. Measuring local $M/L$ values in disk galaxies (assuming conventional dynamics) should give the following results: In regions of the galaxy where $V^2/r \gg a_0$ the local $M/L$ values should show no indication of hidden mass. At a certain transition radius, local $M/L$ should start to increase rapidly. The transition radius should occur where $V^2/r \approx a_0$. This test has the following advantages: (a) It does not require an absolute calibration of $M/L$ as we are concerned only with variations of this quantity; (b) Effects of the modified dynamics manifest themselves more clearly in local mass determination than in the integrated masses; and (c) In many cases this test requires information on local behavior in the disk only while the spheroid can be neglected. This makes the determination of mass from velocity more certain.
6. Disk galaxies with low surface brightness provide particularly strong tests (a study of a sample of such galaxies is described by Strom 1982 and by Romanishin et al. 1982). As low surface brightness means small accelerations, the effects of the modification should be more noticeable in such galaxies. We predict, for example, that the proportionality factor in the $M \propto V_\infty^4$ relation for these galaxies is the same as for the high surface density galaxies. In contrast, if one wants to obtain a correlation $M \propto V_\infty^4$ in the conventional dynamics (with additional assumptions), one is led to the relation $M \propto \Sigma^{-1} V_\infty^4$ (see, for example, Aaronson, Huchra, and Mould 1979), where $\Sigma$ is the average surface brightness. This implies that low surface density galaxies, of a given velocity, have a mass higher than predicted by the $M$-$V$ relation derived for normal surface density galaxies.
We also predict that the lower the average surface density of a galaxy is, the smaller is the transition radius, defined in prediction 5, in units of the galaxy's scale length. In fact, if the average surface density is very small we may have a galaxy in which $V^2/r < a_0$ everywhere, and analysis with conventional dynamics should yield local $M/L$ values starting to increase from very small radii.
7. As the study of model rotation curves shows, we predict a correlation between the value of the average surface density (or brightness) of a galaxy and the steepness with which the rotational velocity rises to its asymptotic value (as measured, for example, by the radius at which $V = V_\infty/2$ in units of the scale length of the disk). Small surface densities imply slow rise of $V$.

IX. DISCUSSION

The main results of this paper can be summarized by the statement that the modified dynamics eliminates the need to assume hidden mass in galaxies. The effects in galaxies which I have considered, and which are commonly attributed to such hidden mass, are readily explained by the modification. More specifically:

# MOND predictions

- The Tully-Fisher Relation
  - Slope = 4
  - Normalization = $1/(a_0 G)$
  - Fundamentally a relation between Disk Mass and $V_{flat}$
  - No Dependence on Surface Brightness

- Dependence of conventional M/L on radius and surface brightness

- Rotation Curve Shapes

- Surface Density ~ Surface Brightness

- Detailed Rotation Curve Fits

- Stellar Population Mass-to-Light Ratios

*"Disk Galaxies with low surface brightness provide particularly strong tests"*
*None of the following data existed in 1983.*
*At that time, LSB galaxies were widely thought not to exist.*

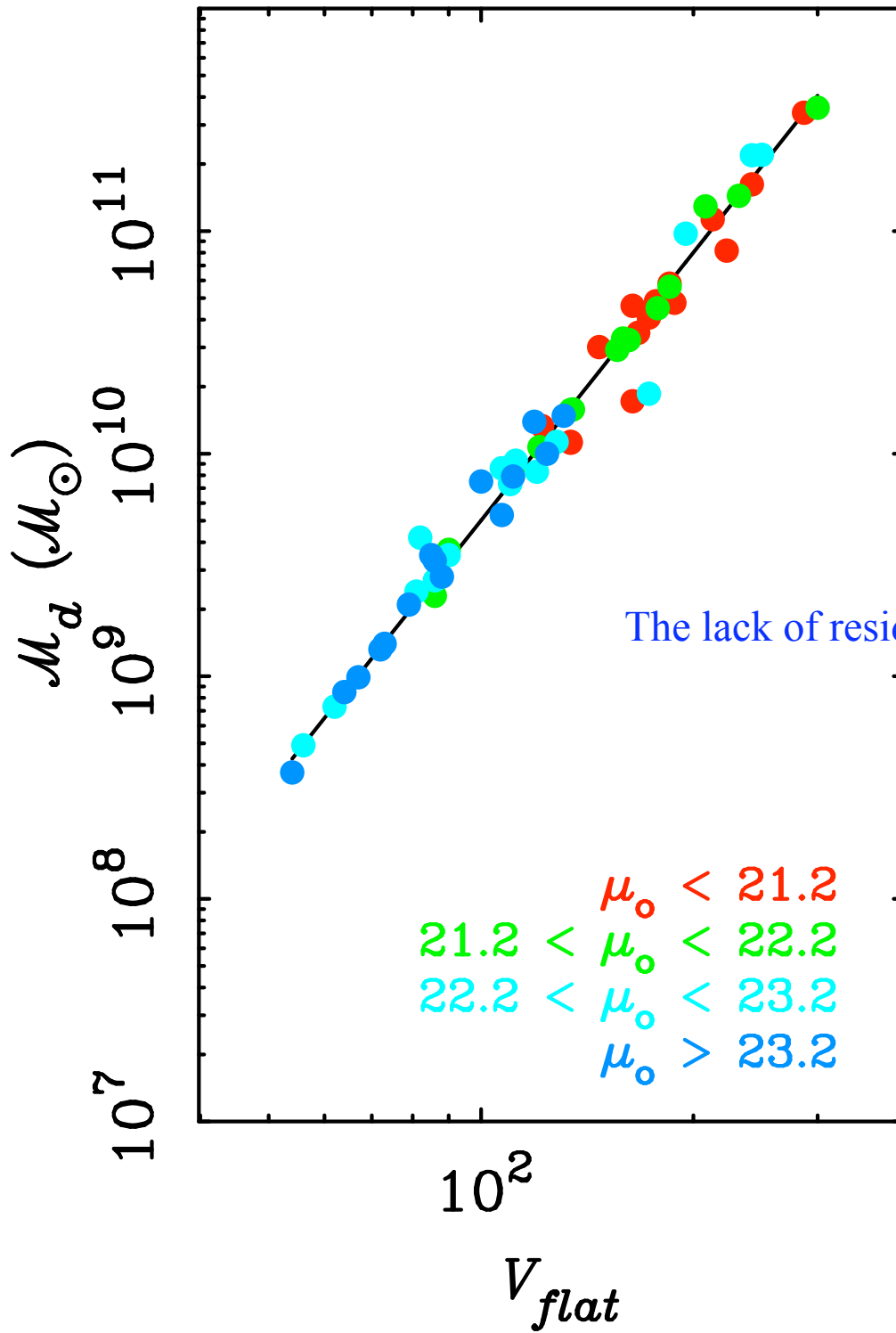

# MOND predictions

- The Tully-Fisher Relation

✔ • Slope = 4
✔ • Normalization = $1/(a_0 G)$
✔ • Fundamentally a relation between Disk Mass and $V_{flat}$
✔ • No Dependence on Surface Brightness ❗

*The lack of residuals from Tully-Fisher an a priori prediction*

- Dependence of conventional M/L on radius and surface brightness

- Rotation Curve Shapes

- Surface Density ~ Surface Brightness

- Detailed Rotation Curve Fits

- Stellar Population Mass-to-Light Ratios

## Test TF slope by extrapolation to very low velocities:

(McGaugh 2005)

**TABLE 5**
**EXTREME DWARF GALAXY DATA**

| Galaxy | $V_f$ (km s$^{-1}$) | $\mathcal{M}_\star$ ($10^6\,\mathcal{M}_\odot$) | $\mathcal{M}_g$ ($10^6\,\mathcal{M}_\odot$) | References |
|---|---|---|---|---|
| ESO215–G?009 | $51^{+8}_{-9}$ | 23 | 714 | 1 |
| UGC 11583[a] | $48^{+3}_{-4}$ | 119 | 36 | 2, 3 |
| NGC 3741 | $44^{+4}_{-2}$ | 25 | 224 | 4 |
| WLM | $38^{+5}_{-5}$ | 31 | 65 | 5 |
| KK98 251 | $36^{+8}_{-4}$ | 12 | 98 | 3 |
| GR 8 | $25^{+5}_{-4}$ | 5 | 14 | 6 |
| Cam B | $20^{+10}_{-13}$ | 3.5 | 6.6 | 7 |
| DDO 210 | $17^{+3}_{-5}$ | 0.9 | 3.6 | 8 |

[a] UGC 11583 is KK98 250.

REFERENCES.—(1) Warren et al. 2004; (2) McGaugh et al. 2001; (3) Begum & Chengalur 2004a; (4) Begum et al. 2005; (5) Jackson et al. 2004; (6) Begum & Chengalur 2003; (7) Begum et al. 2003; (8) Begum & Chengalur 2004b.

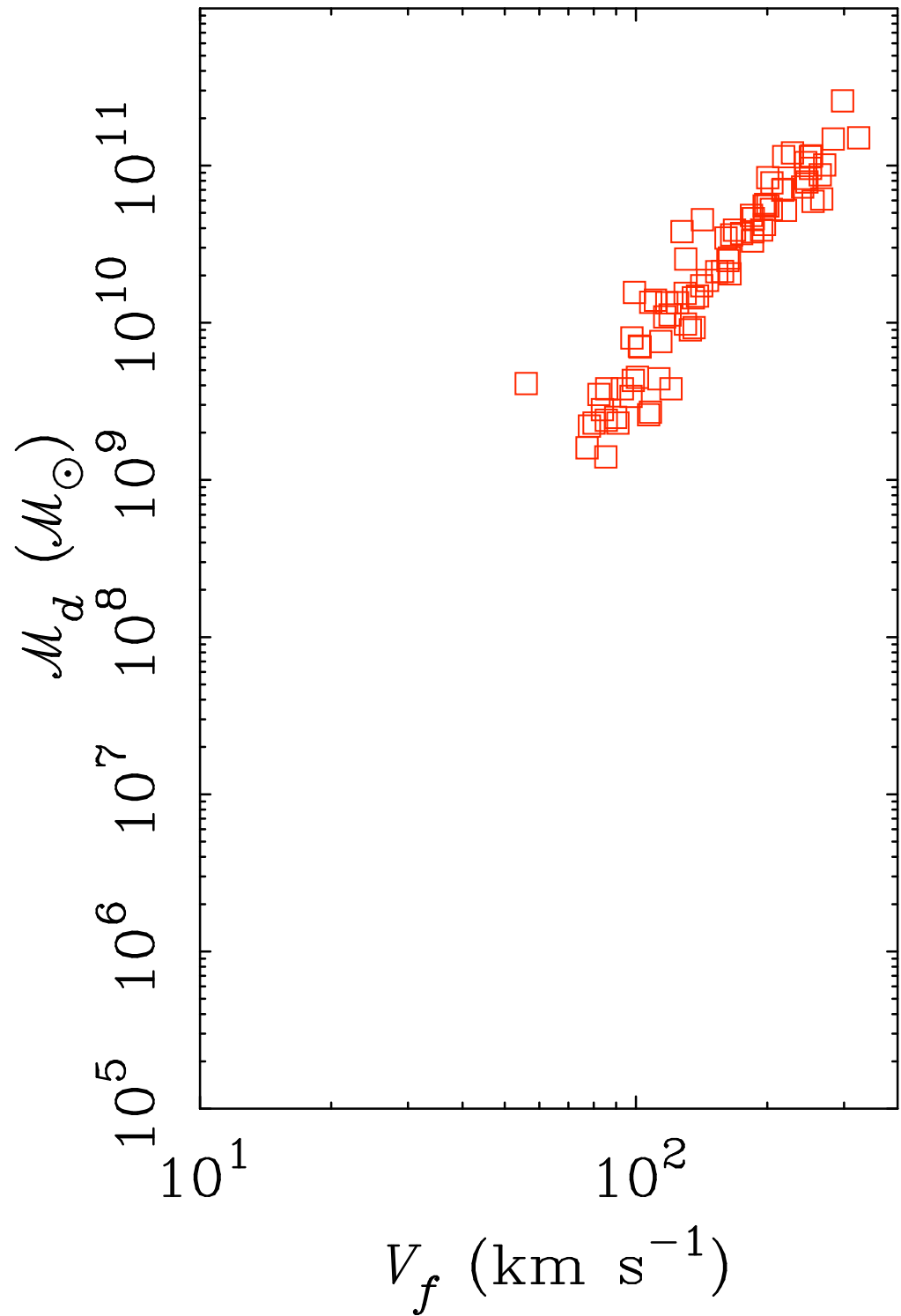

Pizagno et al. (2005)
ApJ, 633, 844

This is typical of the range covered by most Tully-Fisher studies.

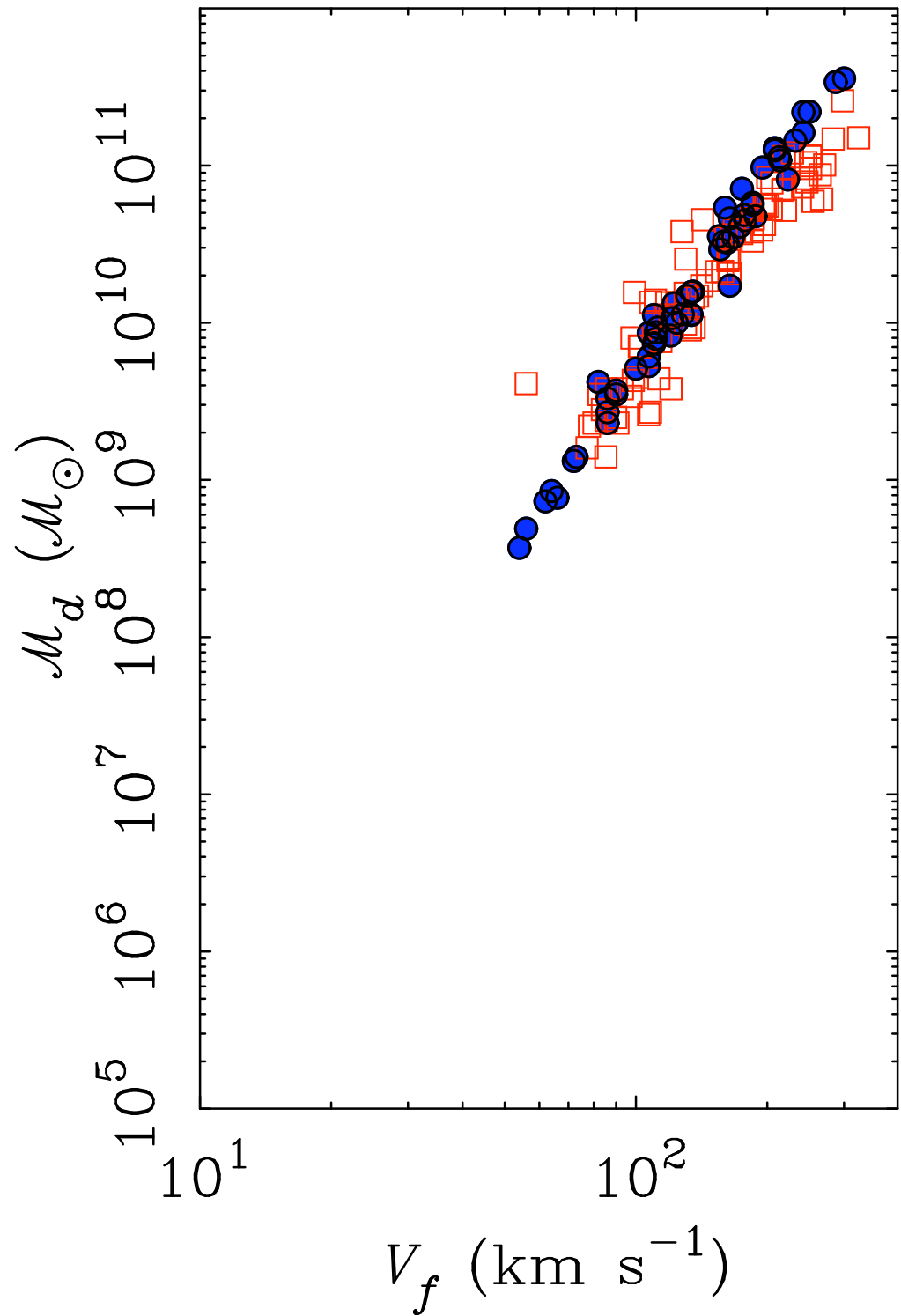

McGaugh (2005)
ApJ, 632, 859

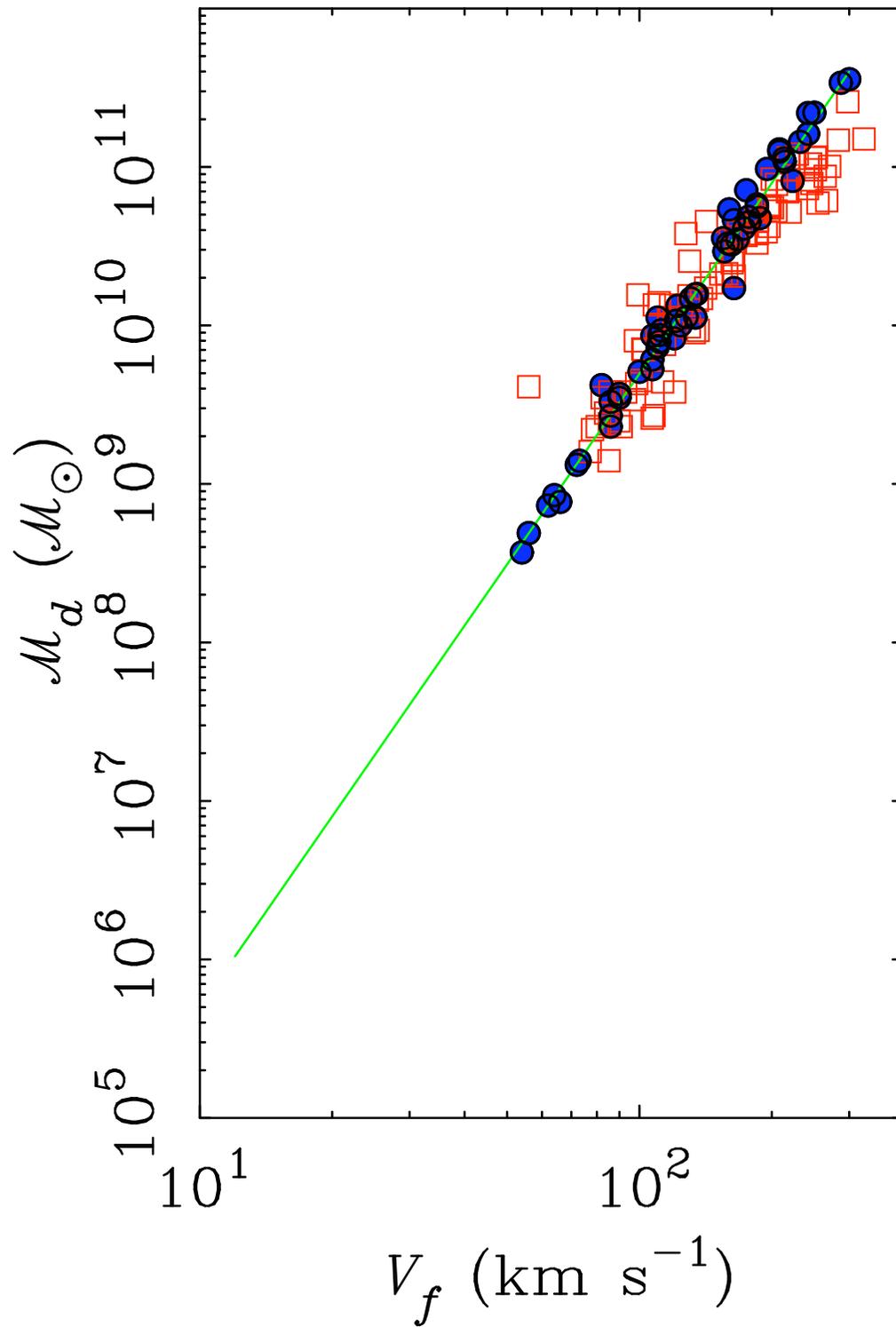

green line:
best fit for mass-to-light ratio estimator with minimum scatter:

$$\mathcal{M}_d = 50 V^4$$

$(\mathcal{M}_\odot)$ $(\text{km s}^{-1})$

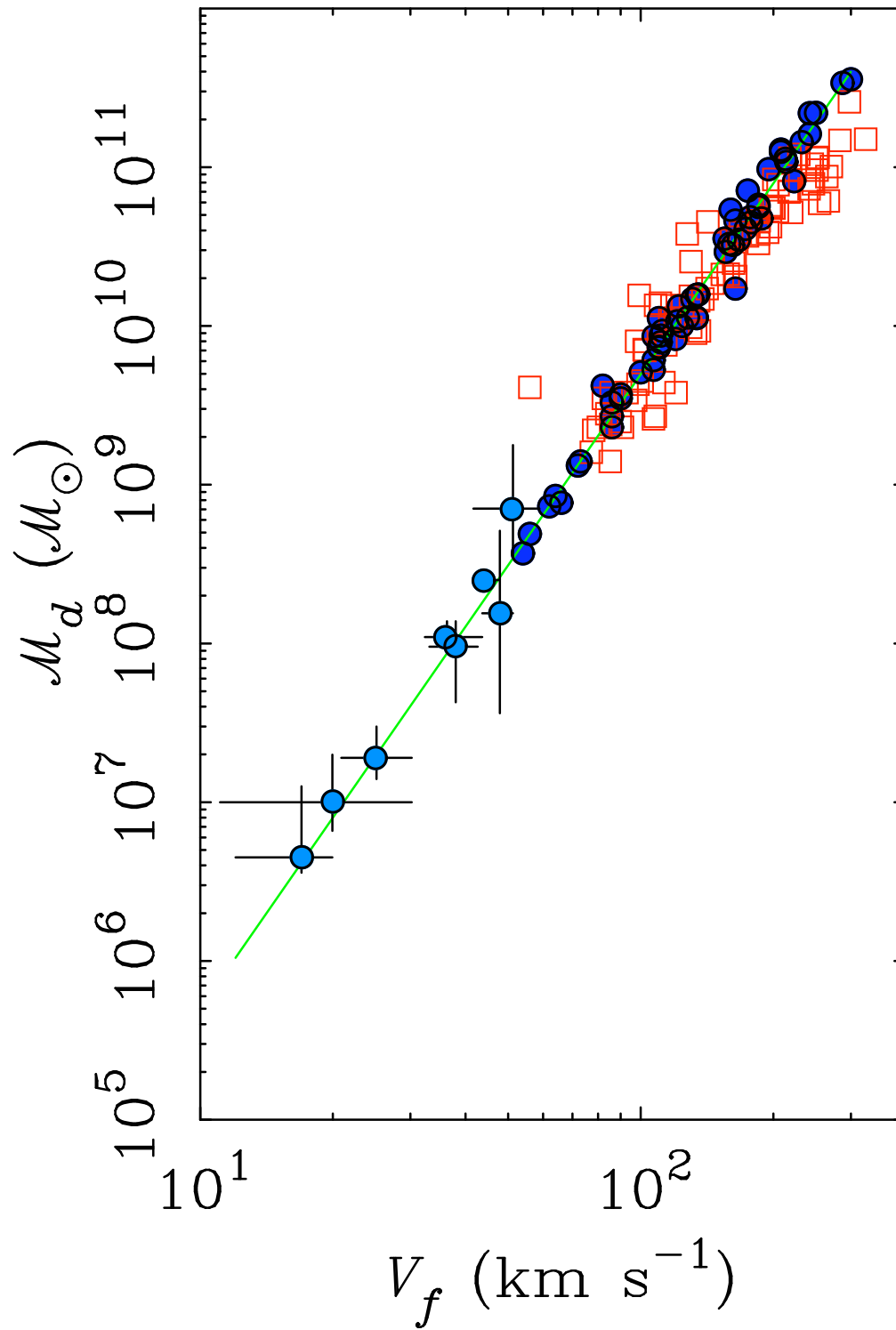

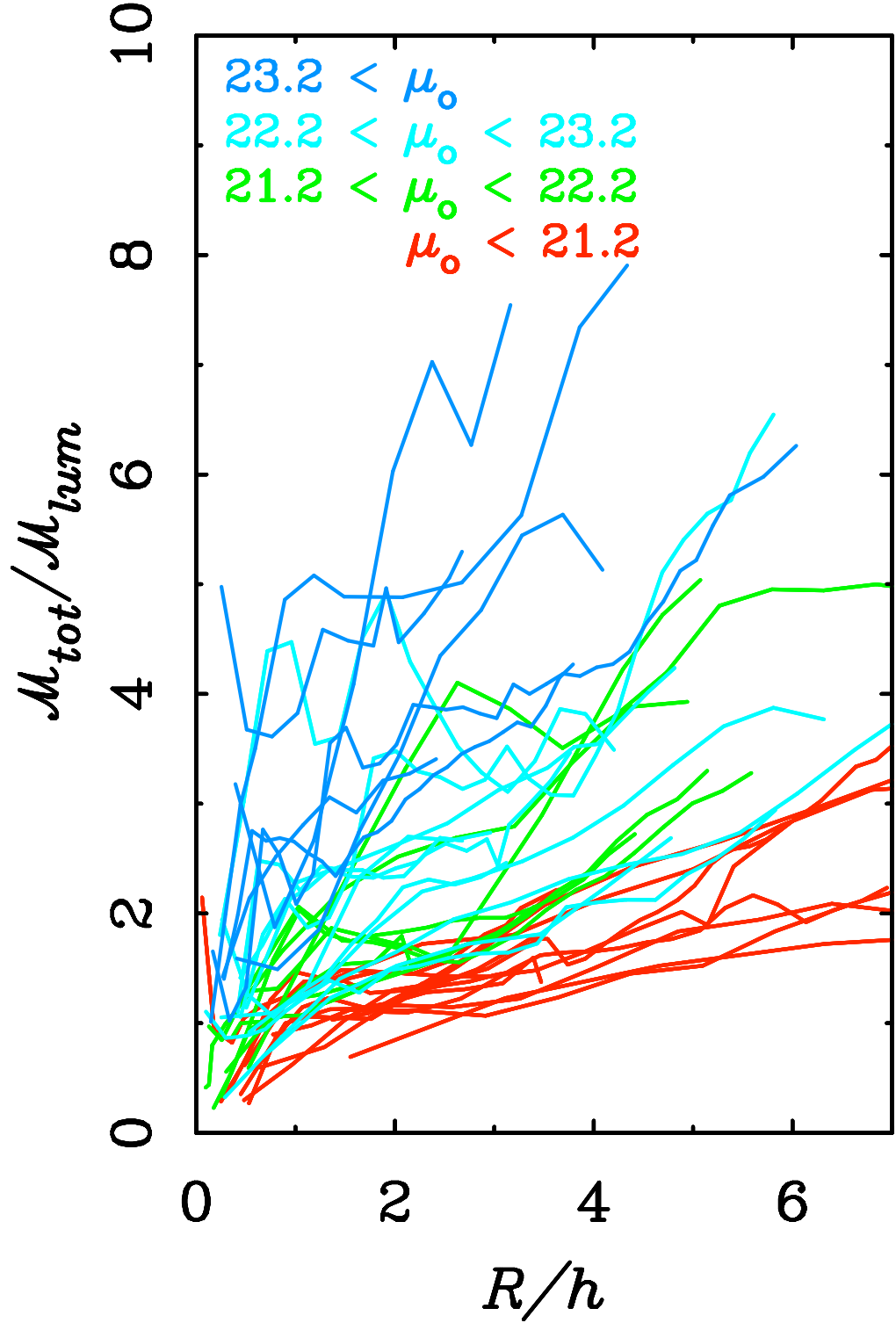

## MOND predictions

- The Tully-Fisher Relation

  ✓ Slope = 4
  ✓ Normalization = $1/(a_0 G)$
  ✓ Fundamentally a relation between Disk Mass and $V_{flat}$
  ✓ No Dependence on Surface Brightness

Other *a priori* MOND predictions

  ✓ Dependence of conventional M/L on radius and surface brightness

- Rotation Curve Shapes

- Surface Density ~ Surface Brightness

- Detailed Rotation Curve Fits

- Stellar Population Mass-to-Light Ratios

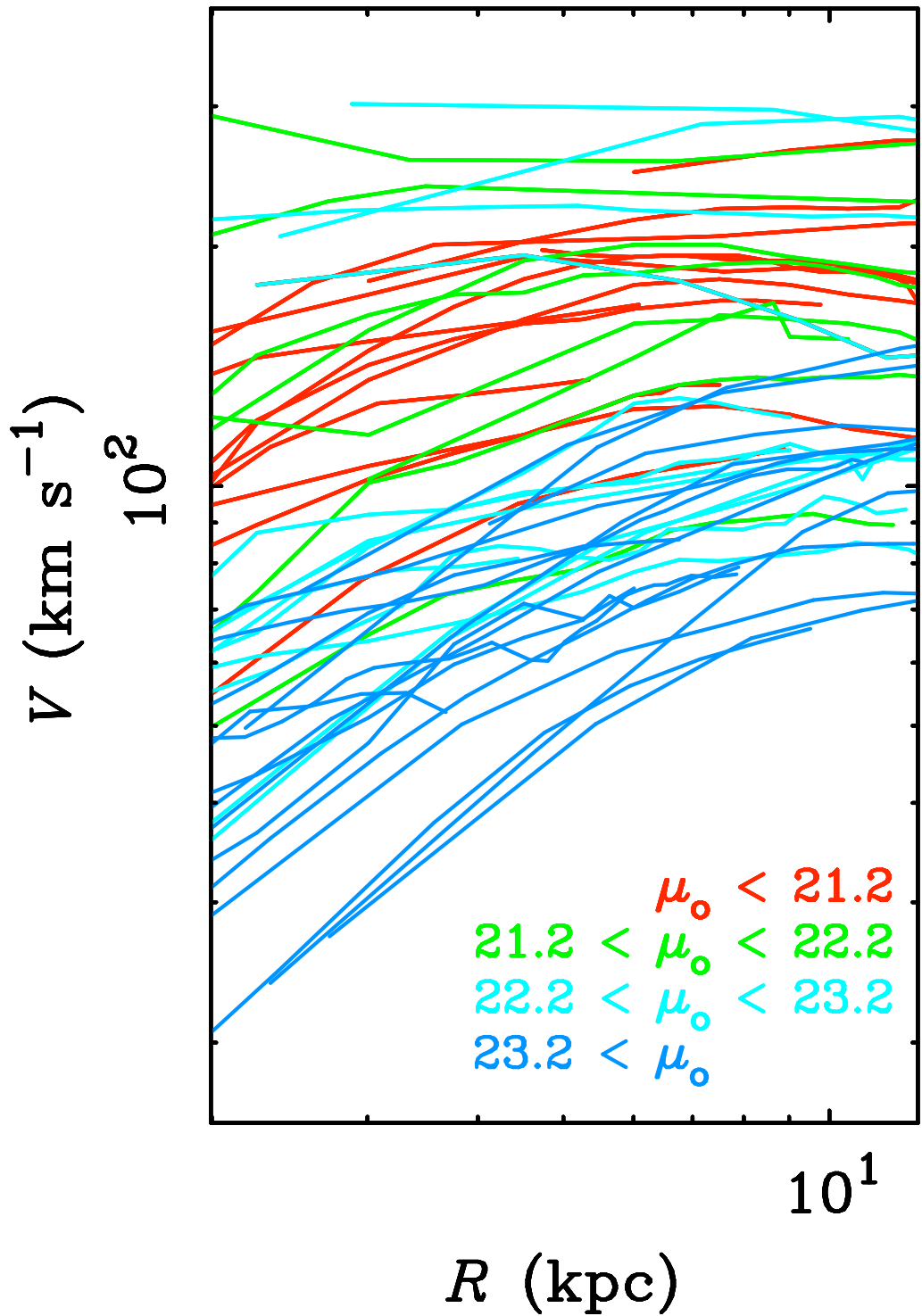

# MOND predictions

- The Tully-Fisher Relation

  ✔ Slope = 4
  ✔ Normalization = $1/(a_0 G)$
  ✔ Fundamentally a relation between Disk Mass and $V_{flat}$
  ✔ No Dependence on Surface Brightness

✔ Dependence of conventional M/L on radius and surface brightness

✔ Rotation Curve Shapes

- Surface Density ~ Surface Brightness

- Detailed Rotation Curve Fits

- Stellar Population Mass-to-Light Ratios

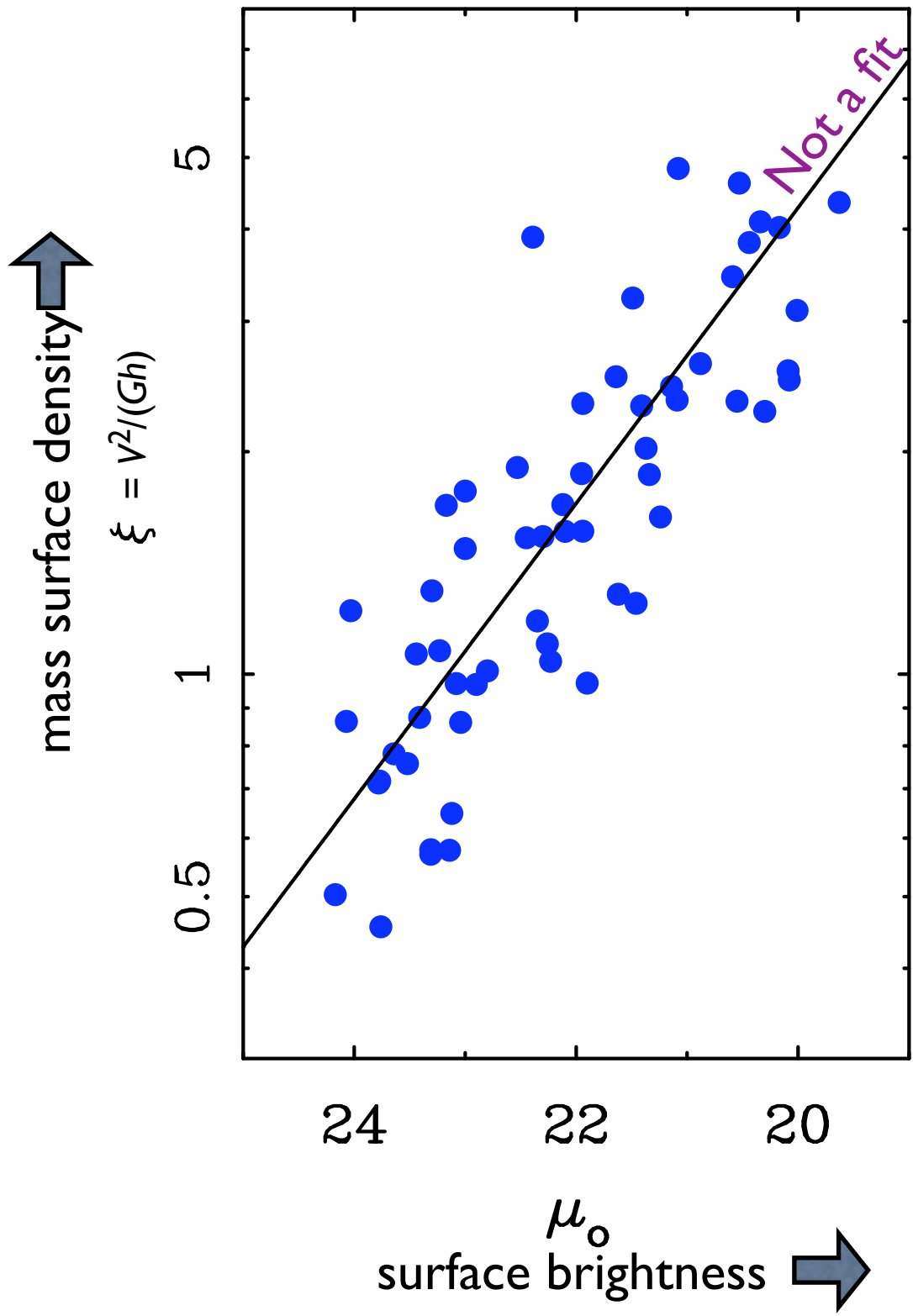

# MOND predictions

- The Tully-Fisher Relation

  ✔ Slope = 4
  ✔ Normalization = $1/(a_0 G)$
  ✔ Fundamentally a relation between Disk Mass and $V_{flat}$
  ✔ No Dependence on Surface Brightness

✔ Dependence of conventional M/L on radius and surface brightness

✔ Rotation Curve Shapes

✔ Surface Density ~ Surface Brightness
No dark matter, so these better correlate

- Detailed Rotation Curve Fits

- Stellar Population Mass-to-Light Ratios

Figure axes: mass surface density $\xi = v^2/(Gh)$ vs $\mu_o$ surface brightness. "Not a fit"

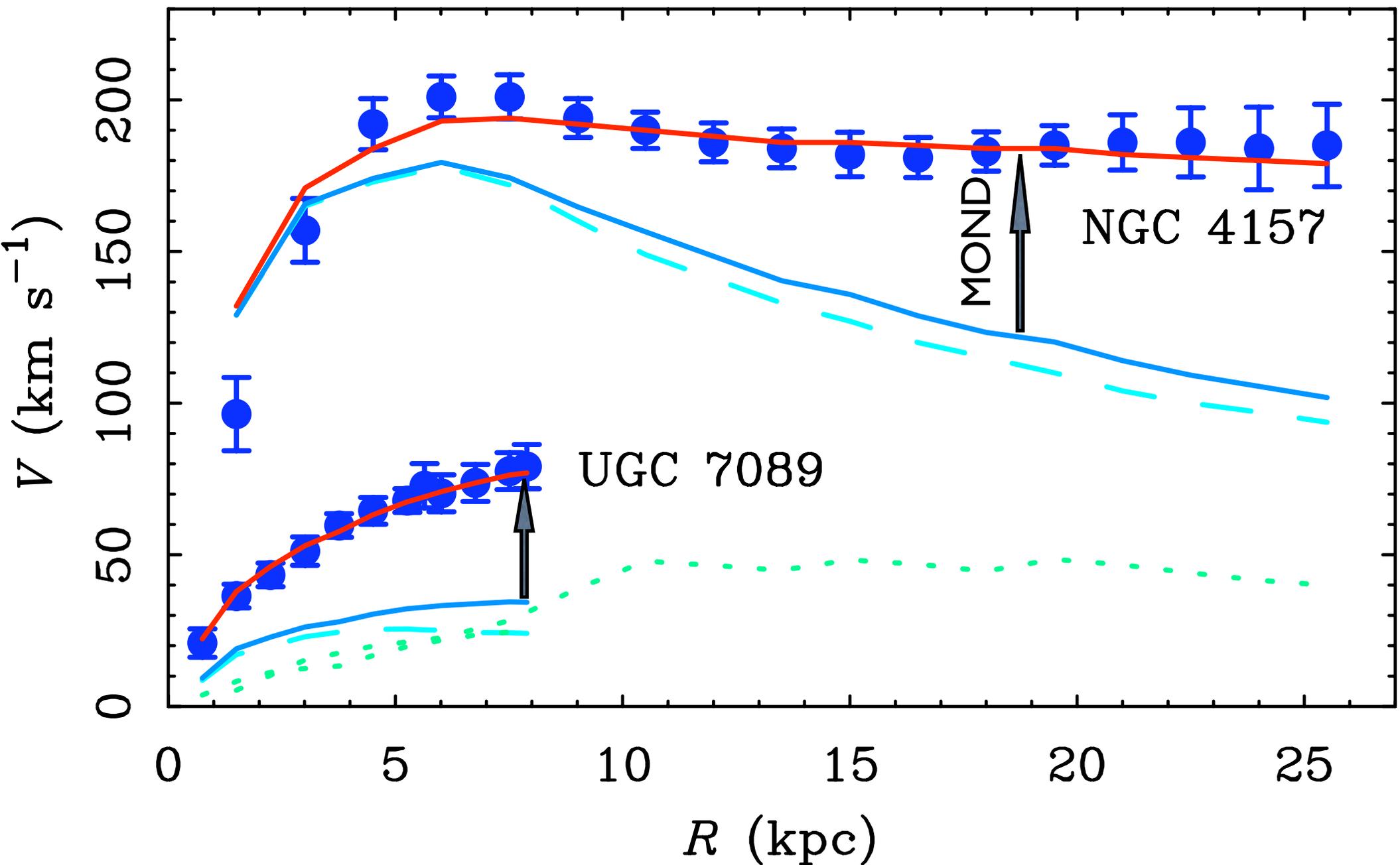

In essence, MOND is a phenomenological scaling relation that maps the Newtonian baryonic rotation curve to the observed one with only a single free parameter, the stellar mass-to-light ratio.

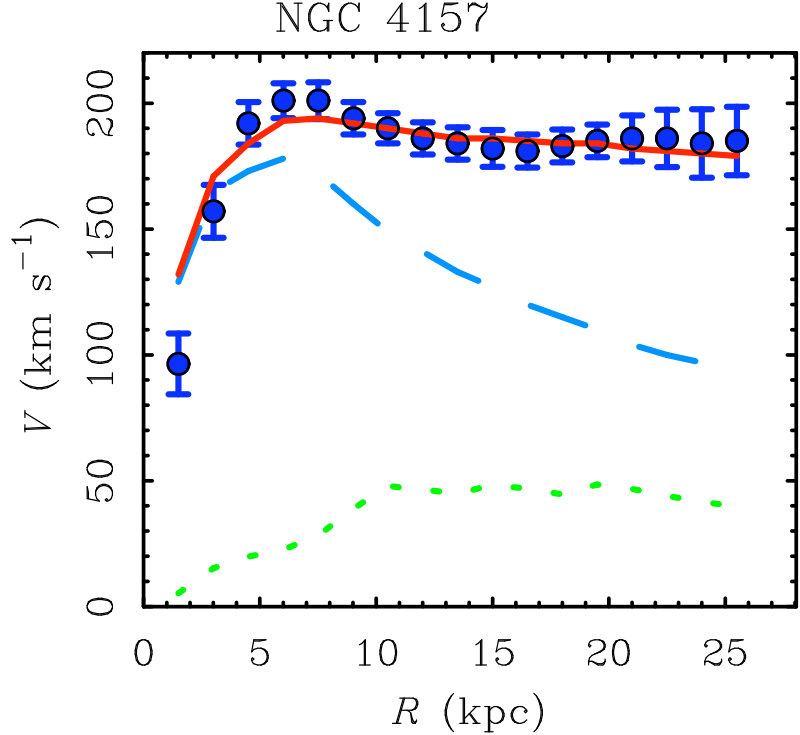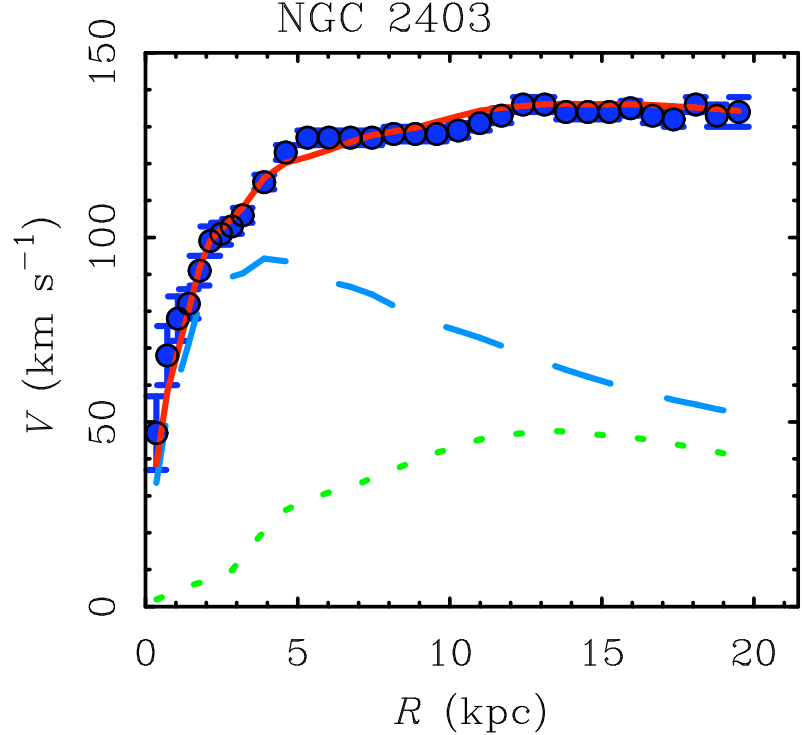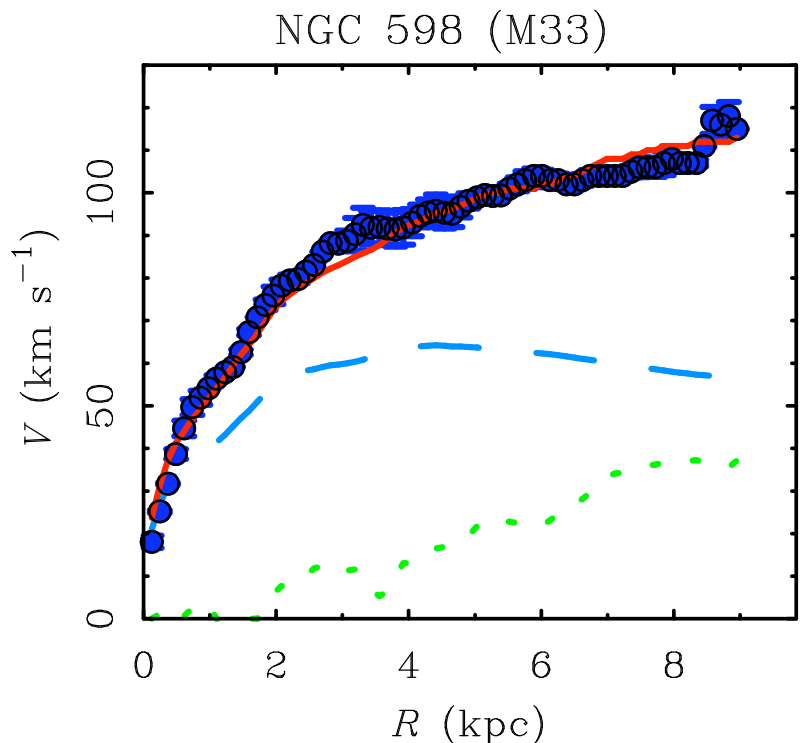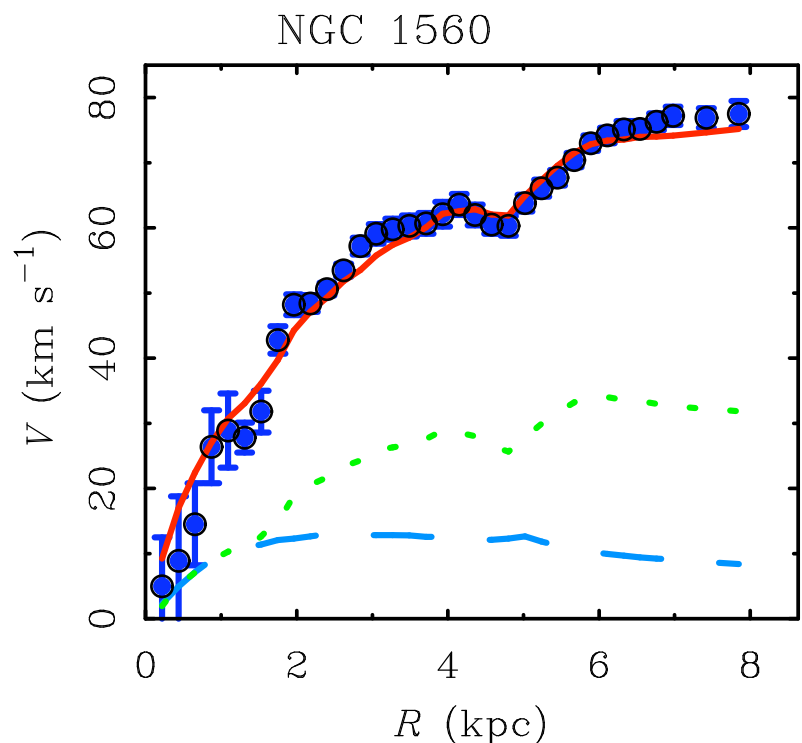

A detailed look at four galaxies.

(Sanders & McGaugh 2002)

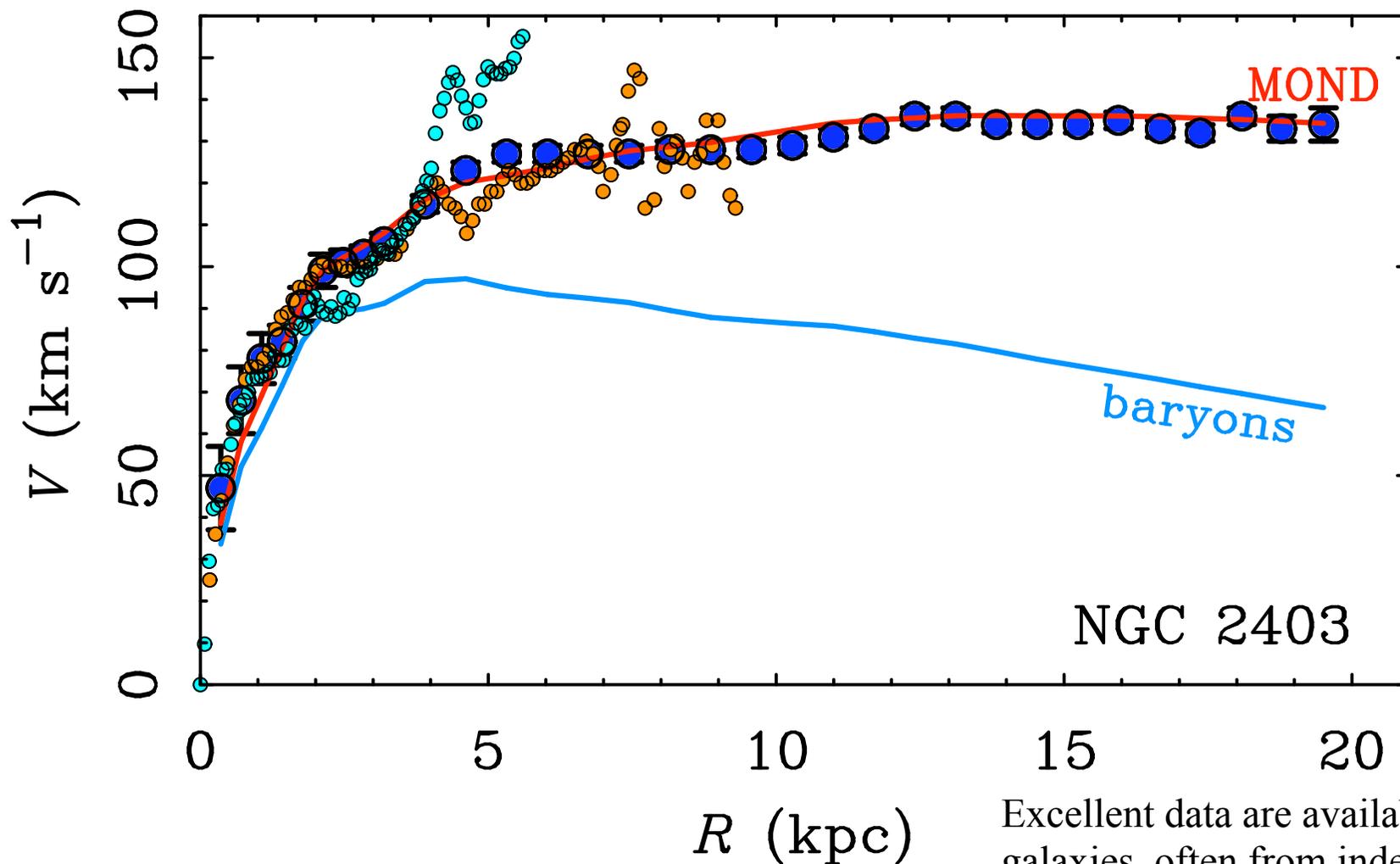

Excellent data are available for some galaxies, often from independent data sets. The HI generally traces further out; Hα generally has better resolution. In this case, Hα goes silly when only one side traces the rotation. All three data sets trace the kink at 3 kpc, which is reflected in the photometry: recall Renzo's rule.

# predictive power: zero free parameters

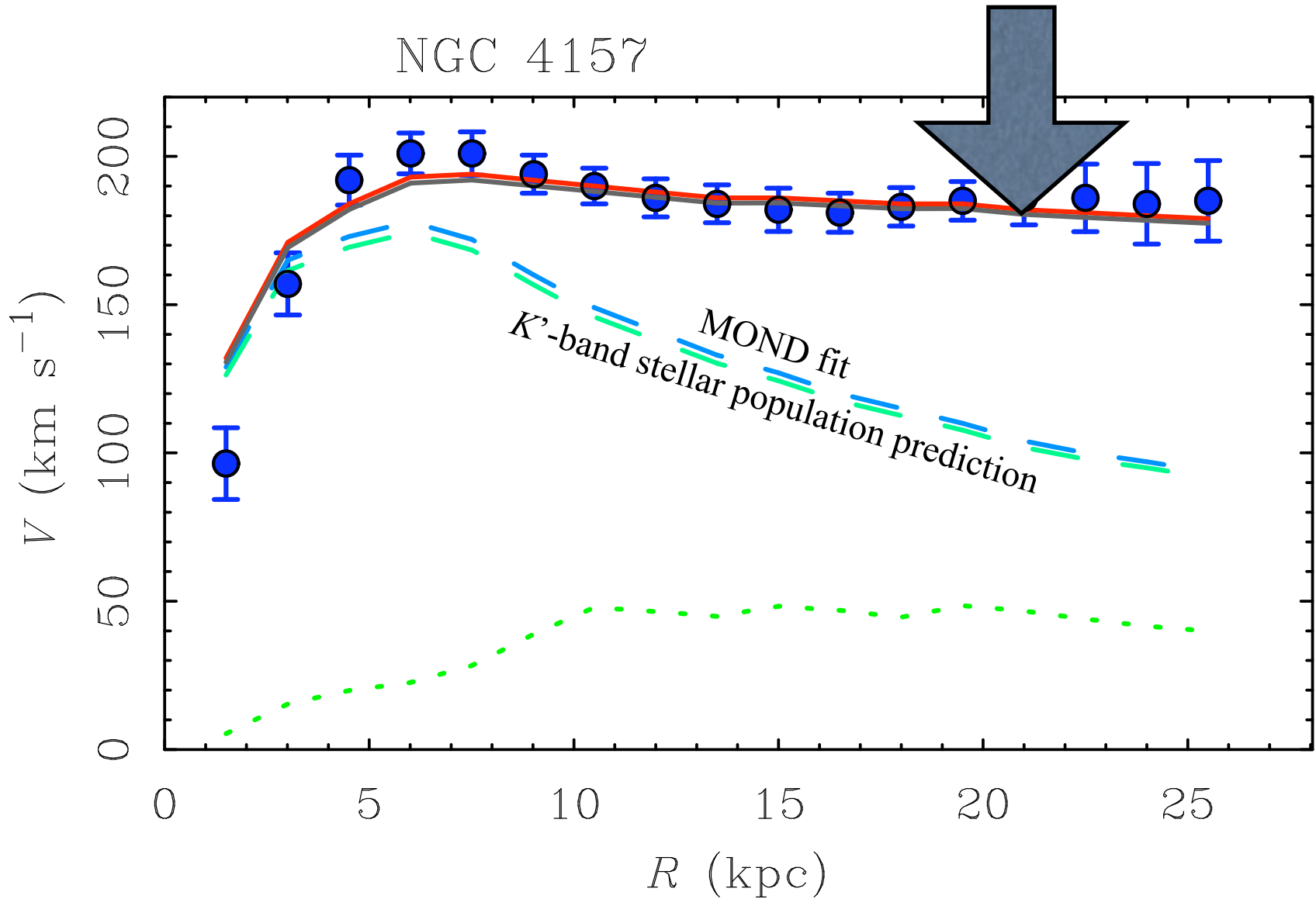

The K'-band is perhaps the best tracer of stellar mass available. One can use stellar population synthesis models (e.g., Bell et al. 2003) to predict the mass-to-light ratio. This does a good job of predicting the mass-to-light ratio needed in MOND fits. In effect, one can predict the rotation curve completely from the photometry.

# M33

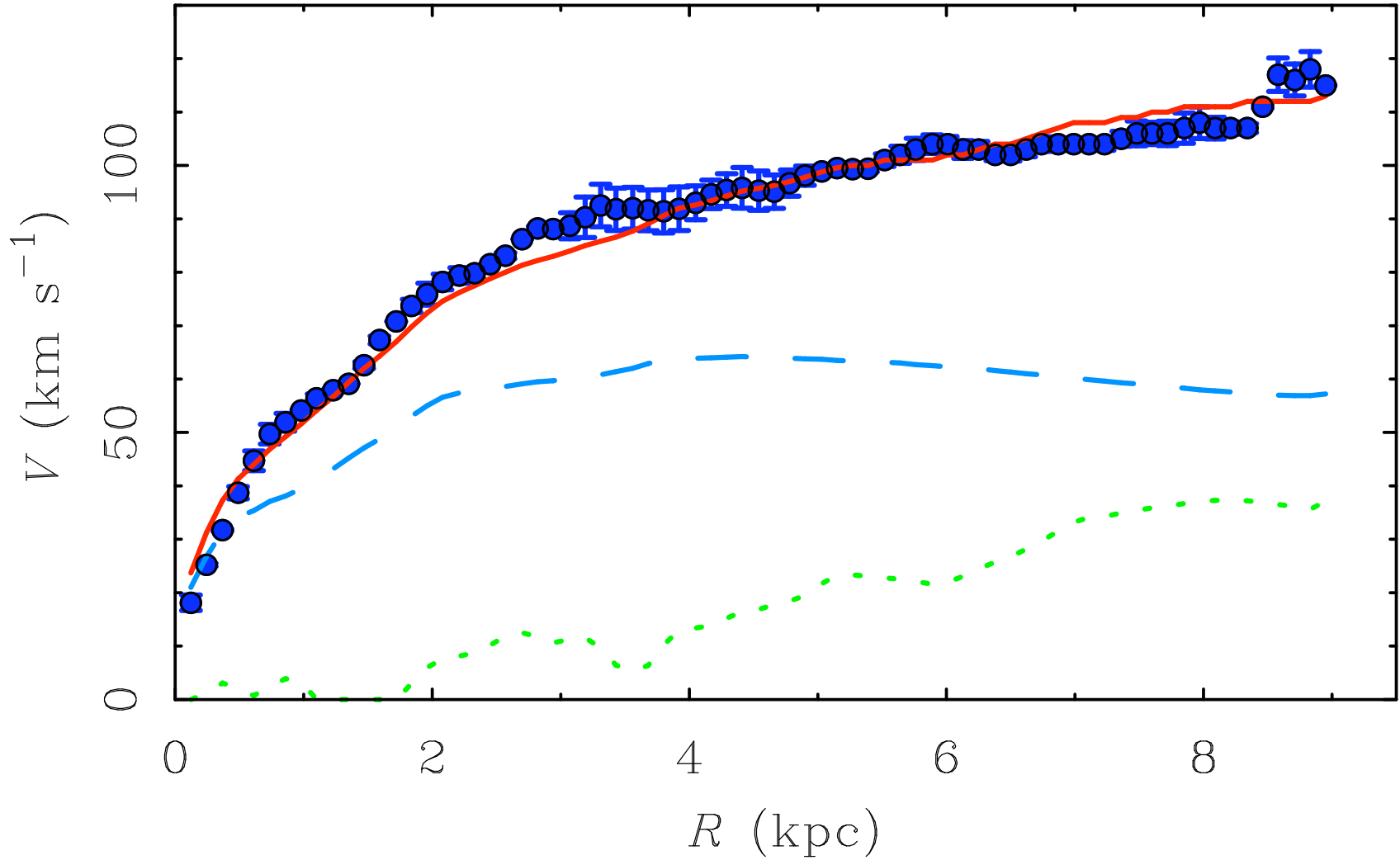

Sanders (1996)

Color gradients should correspond to gradients in the mass-to-light ratio.

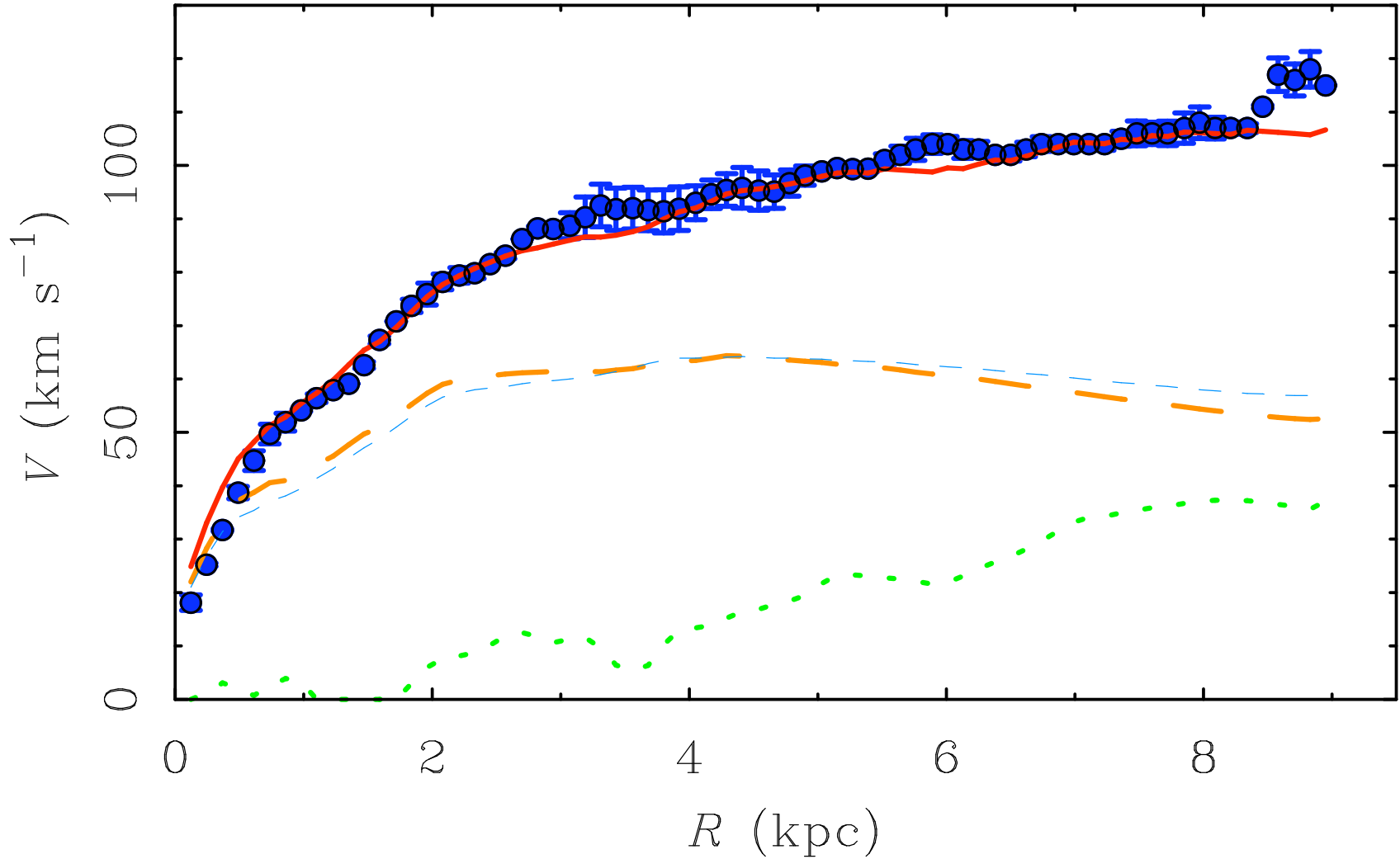

This example shows the effect of correcting for the observed gradient.

Begeman, Broeils, & Sanders (1991)

# NGC 1560

This case is interesting for the prominent kink at large radii, where the quasi-spherical dark matter halo should dominate. (Recall Renzo's rule.)

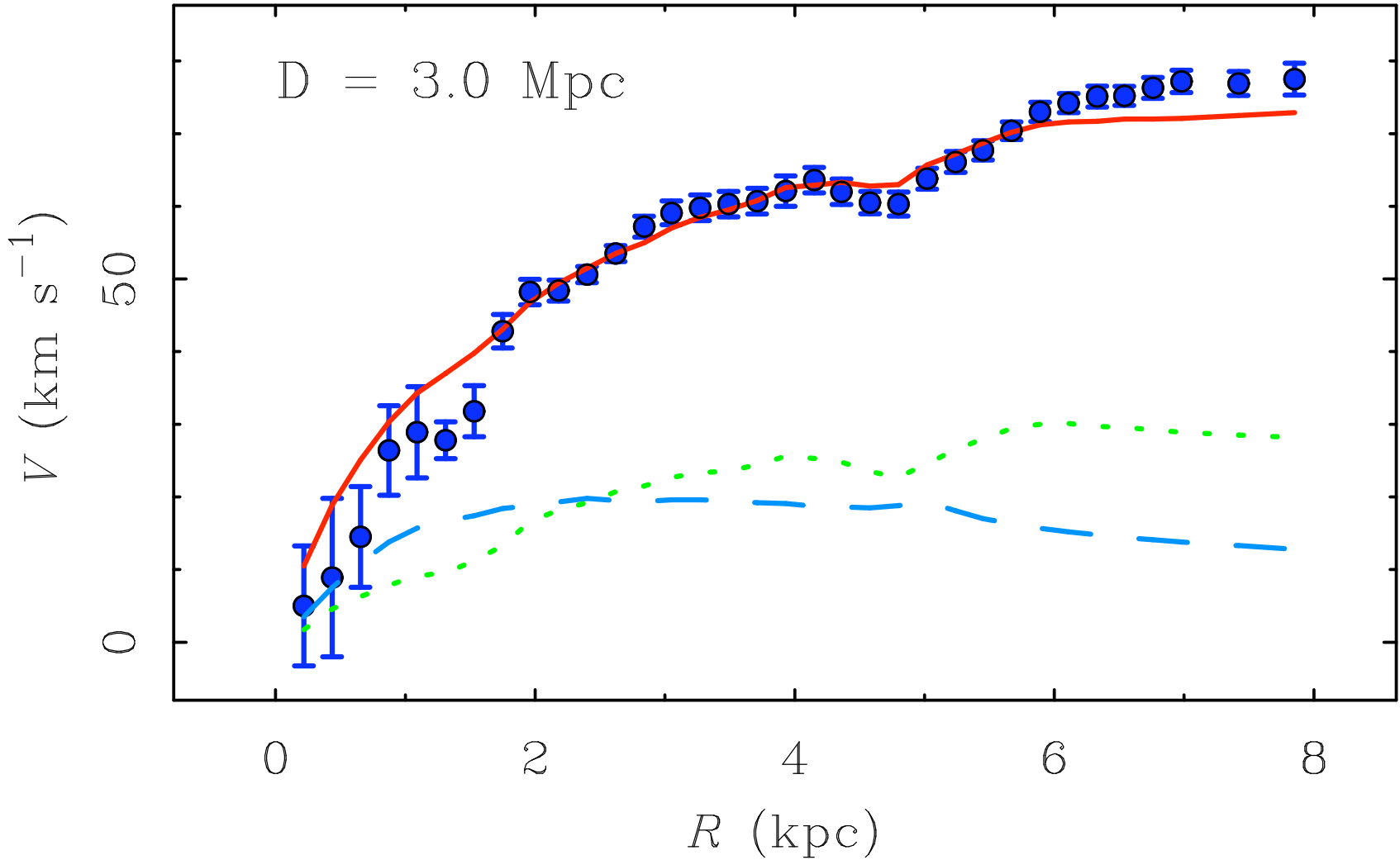

$\Upsilon_* = 0.97$

Begeman, Broeils, & Sanders (1991)

## NGC 1560

Begeman et al. found a better fit if they increased the distance from the value of 3.0 Mpc estimated in 1991 to 3.4 Mpc.

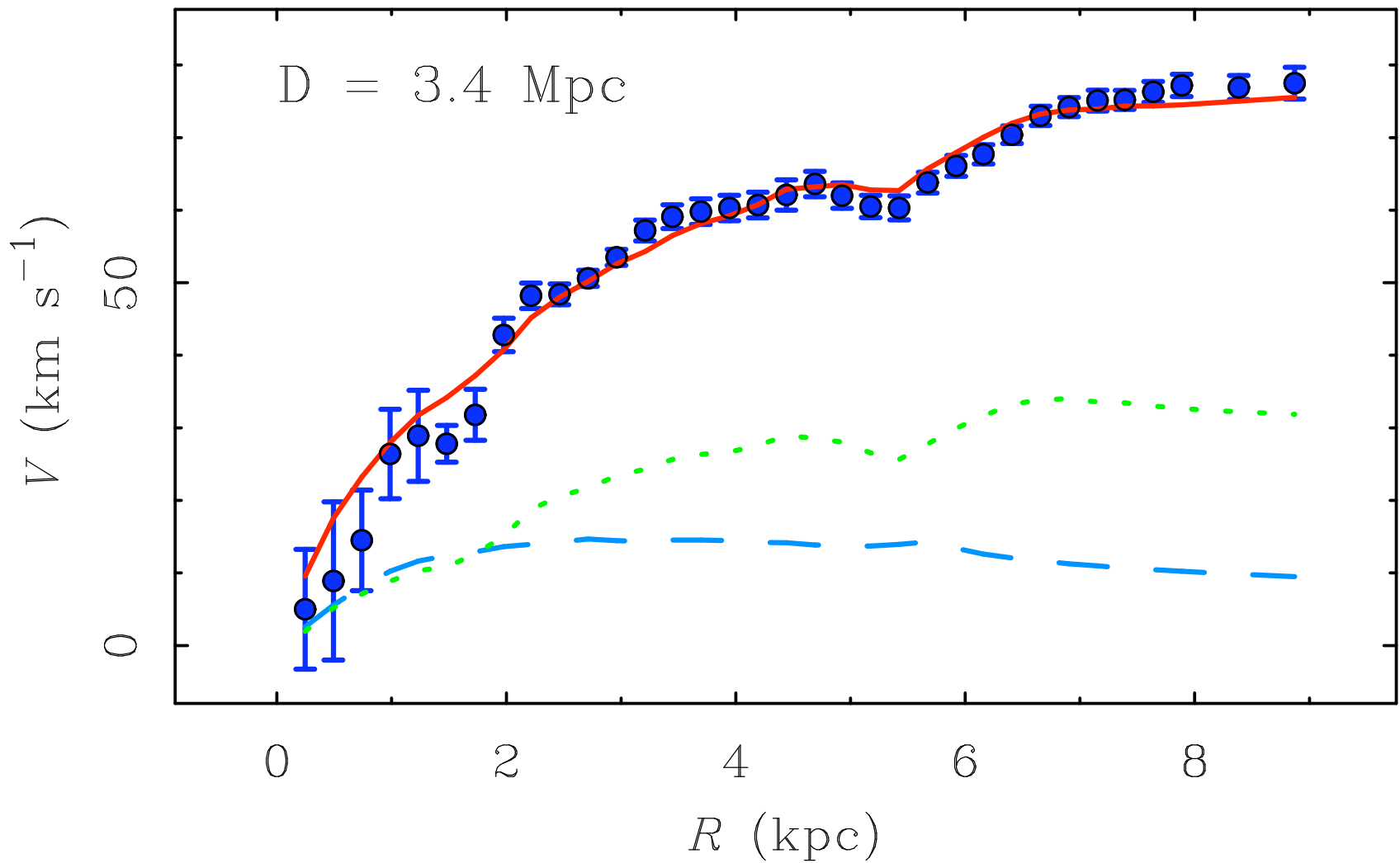

The modern D = 3.45 Mpc, as measured by the Tip of the Red Giant Branch method (Karachentsev *et al.* 2004)

$\Upsilon_* = 0.44$

# Every rotation curve provides a strong test of MOND

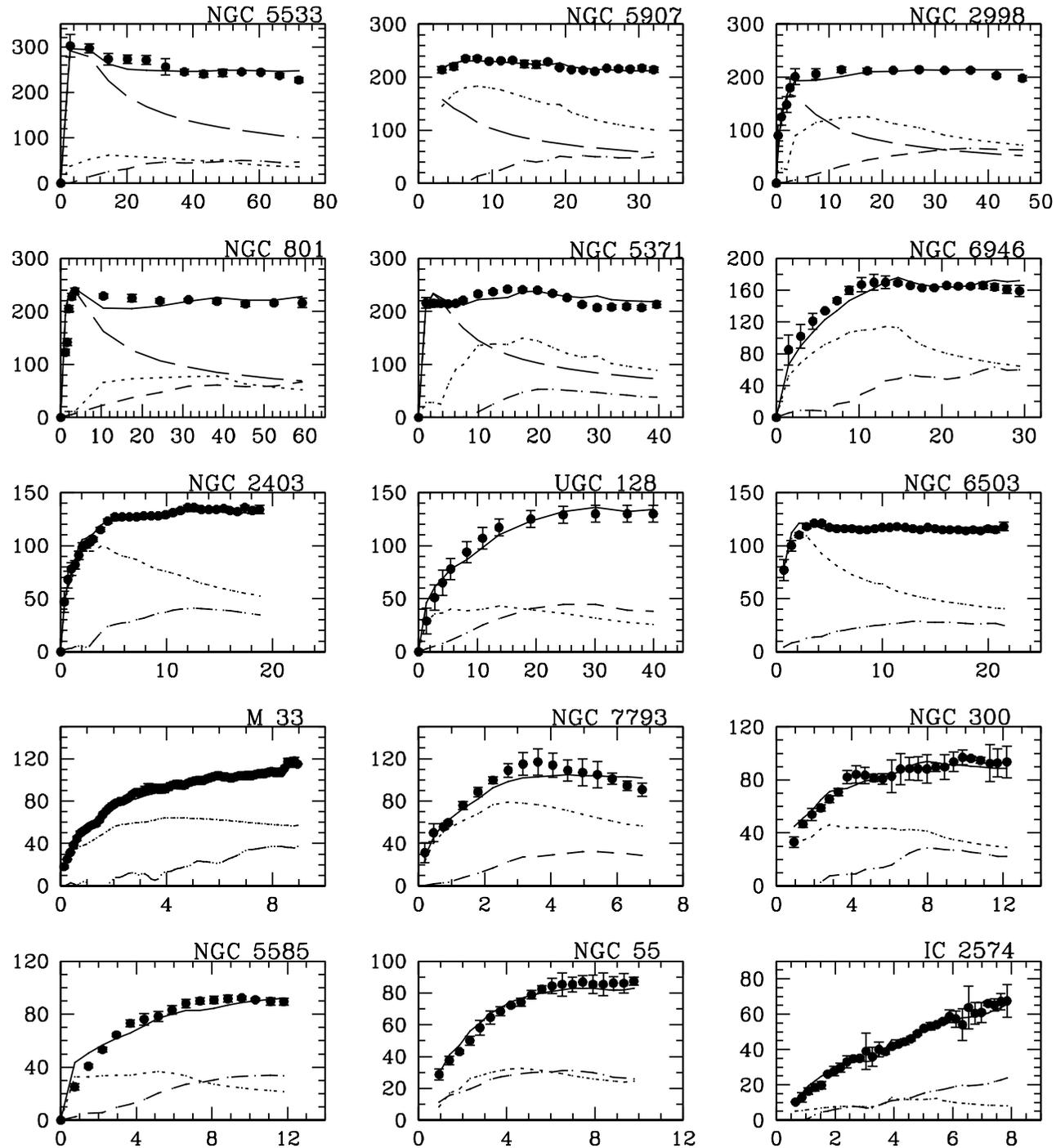

which can be applied to many

Sanders & McGaugh 2002, *ARA&A*, **40**, 263

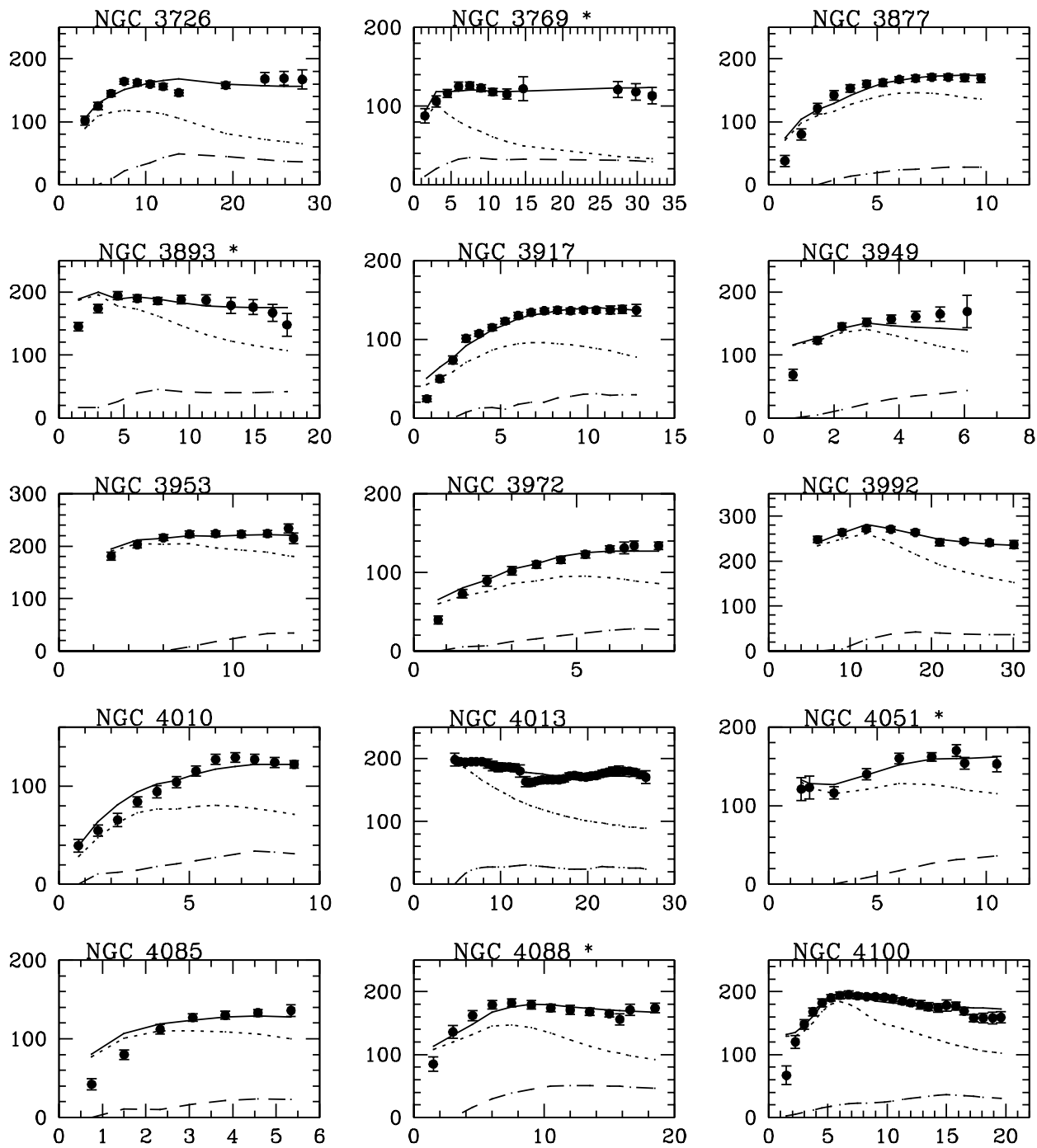

many

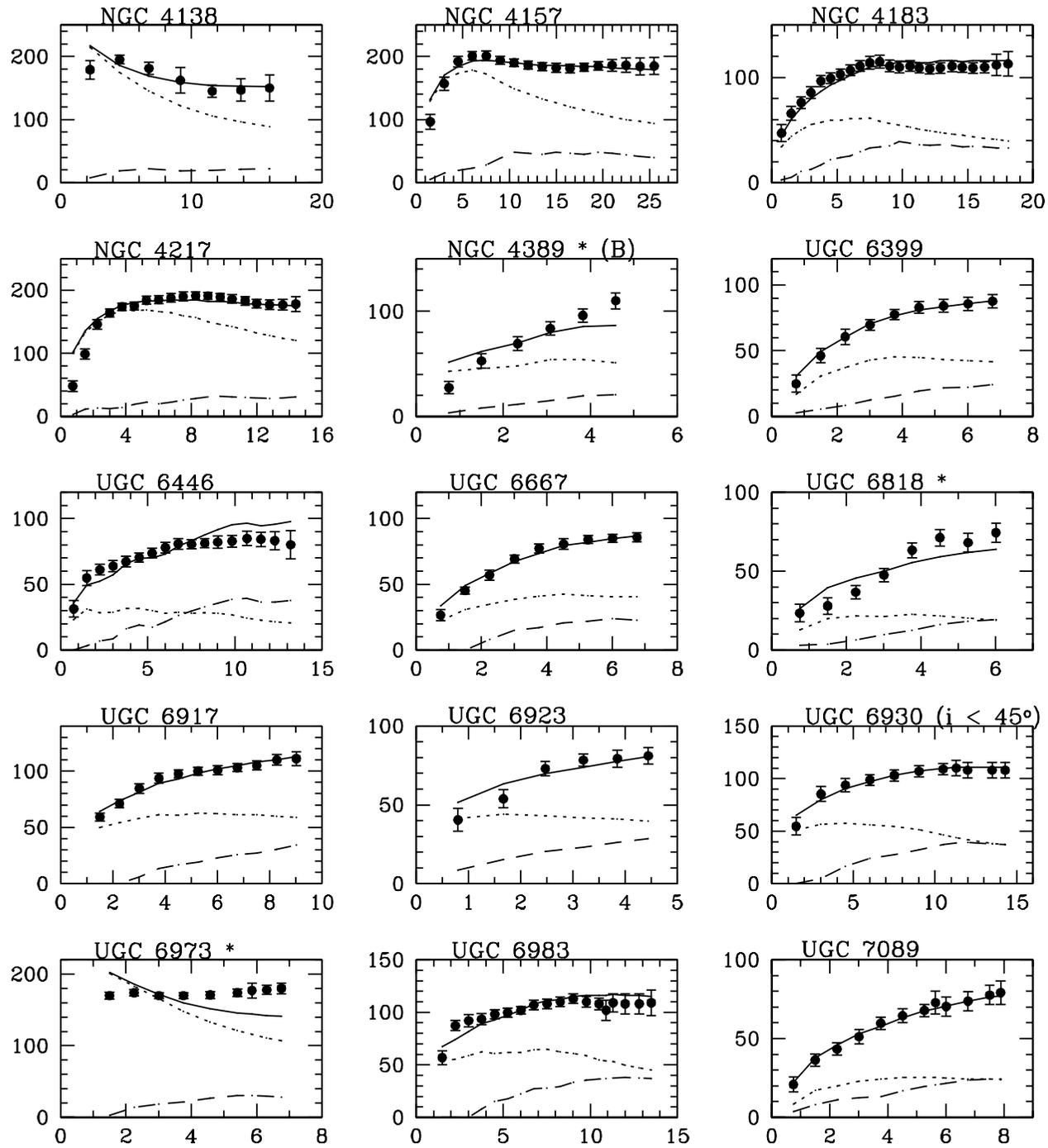

many cases.

# Residuals of MOND fits

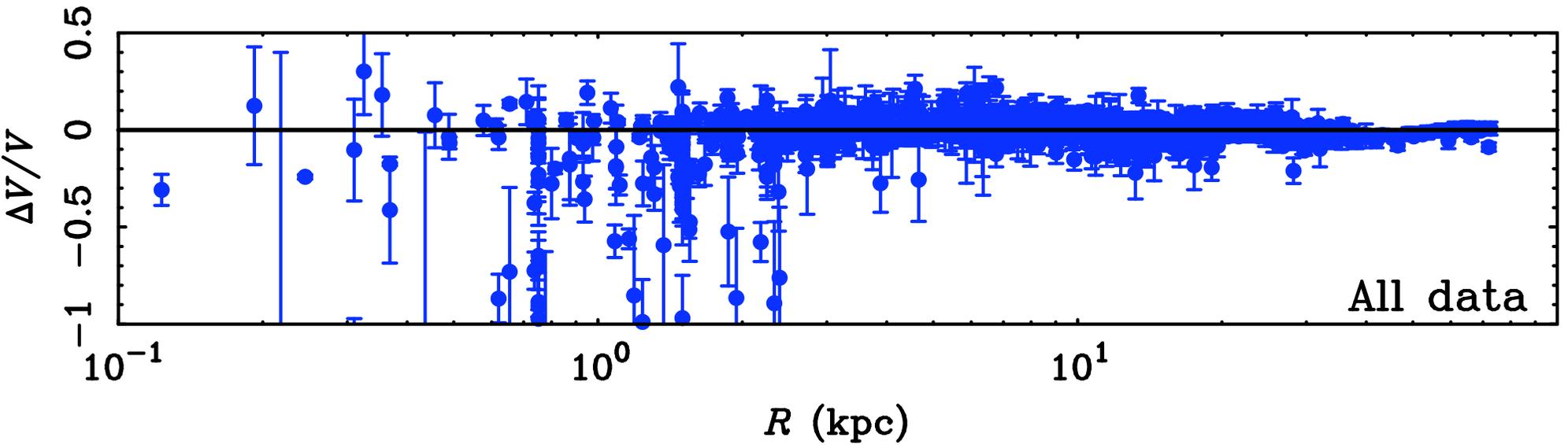

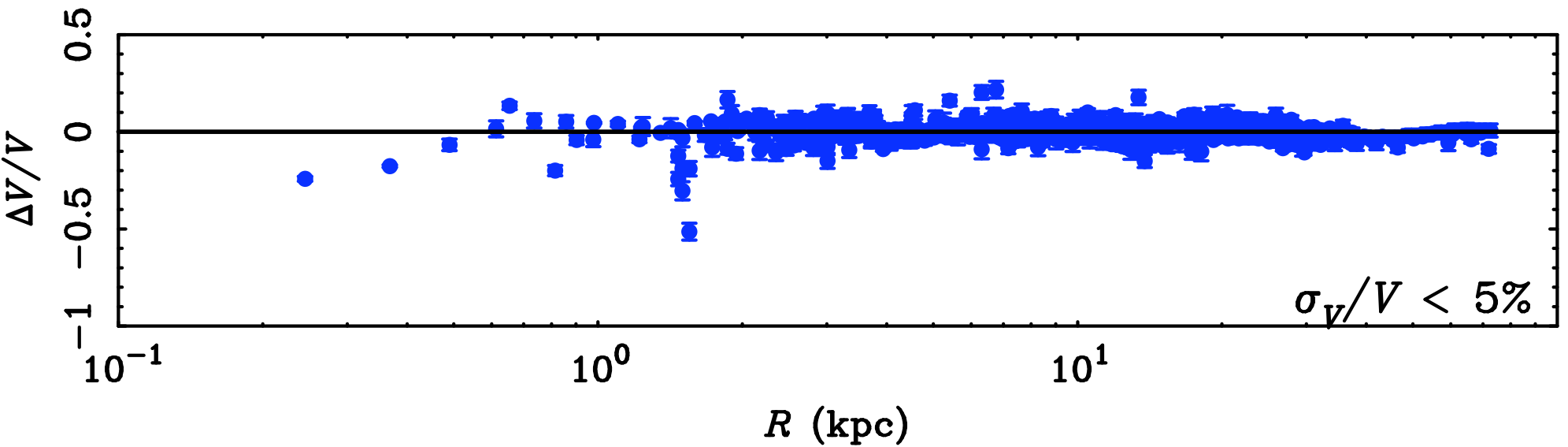

Taken together, the fits are pretty good. There is a slight hint of non-circular motion at small radii, though this is considerably less than commonly invoked for cuspy halos.

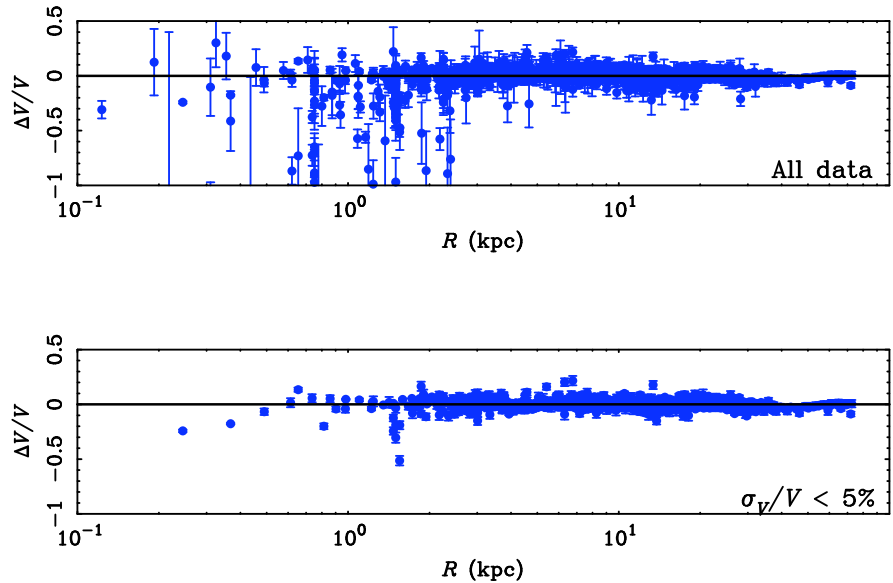

# MOND predictions

- The Tully-Fisher Relation

  ✔ Slope = 4
  ✔ Normalization = $1/(a_0 G)$
  ✔ Fundamentally a relation between Disk Mass and $V_{flat}$
  ✔ No Dependence on Surface Brightness

✔ Dependence of conventional M/L on radius and surface brightness

✔ Rotation Curve Shapes

✔ Surface Density ~ Surface Brightness

✔ Detailed Rotation Curve Fits

- Stellar Population Mass-to-Light Ratios

The success of detailed rotation curve fits is highly non-trivial. Once the form of the force law is specified, the dynamics must follow from the observed baryon distribution.

This procedure is much, much, much more strongly constrained than fits with an invisible dark matter component, which can be arranged however needed. Such fits have a minimum of three free parameters, resulting in enormous freedom and numerous degeneracies.

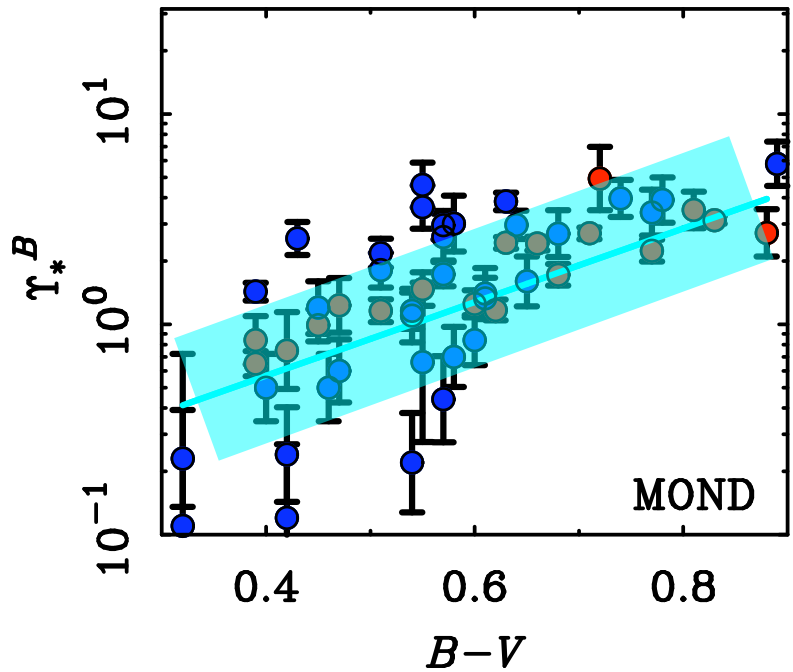

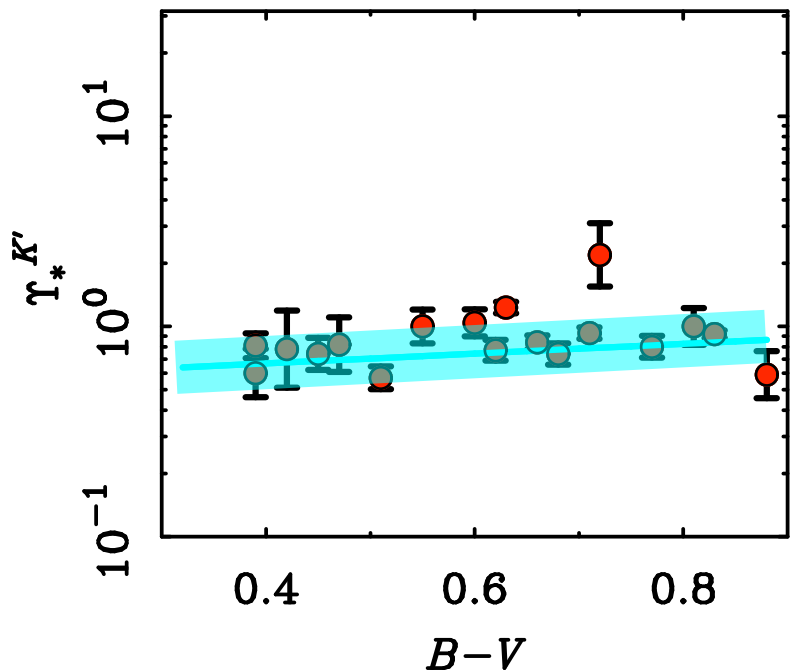

The single free parameter of a MOND fit is the stellar mass-to-light ratio. This is subject to an independent check against the expectations of stellar population synthesis models.

These compare favorably (lines from Bell et al. 2003). MOND fit mass-to-light ratios reproduce not only the mean expected value, but also the trend expected with color and the smaller scatter expected in redder bands.

# MOND predictions

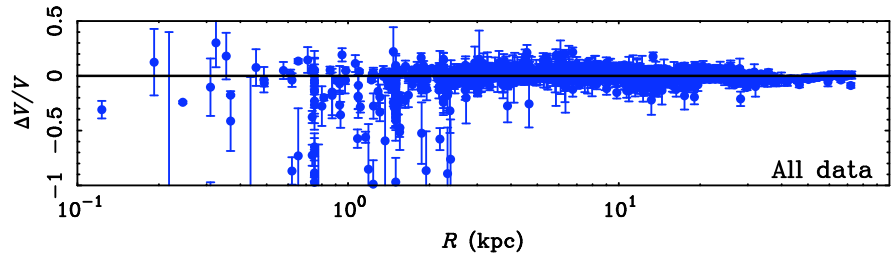

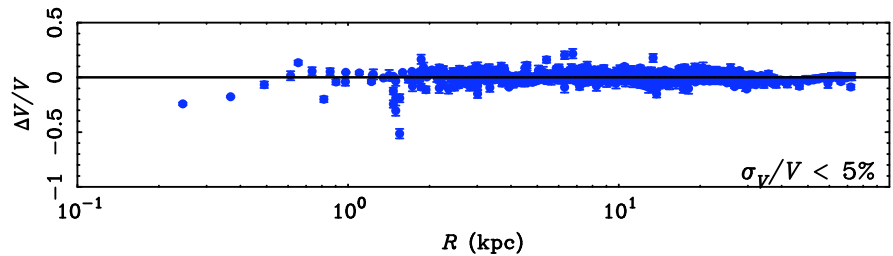

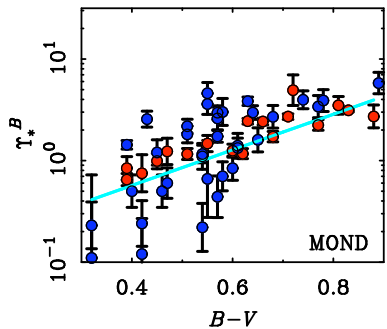

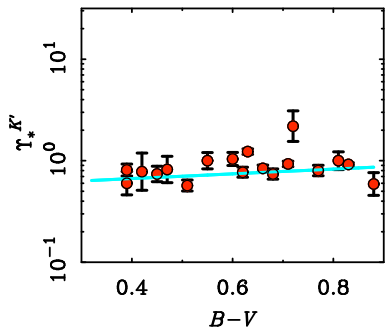

- The Tully-Fisher Relation

  ✔ Slope = 4
  ✔ Normalization = $1/(a_0 G)$
  ✔ Fundamentally a relation between Disk Mass and $V_{flat}$
  ✔ No Dependence on Surface Brightness

  ✔ Dependence of conventional M/L on radius and surface brightness

  ✔ Rotation Curve Shapes

  ✔ Surface Density ~ Surface Brightness

  ✔ Detailed Rotation Curve Fits

  ✔ Stellar Population Mass-to-Light Ratios

What are we suppose to conclude from this? That MOND is wrong?

There is a considerable amount of phenomenological information beyond rotation curves.

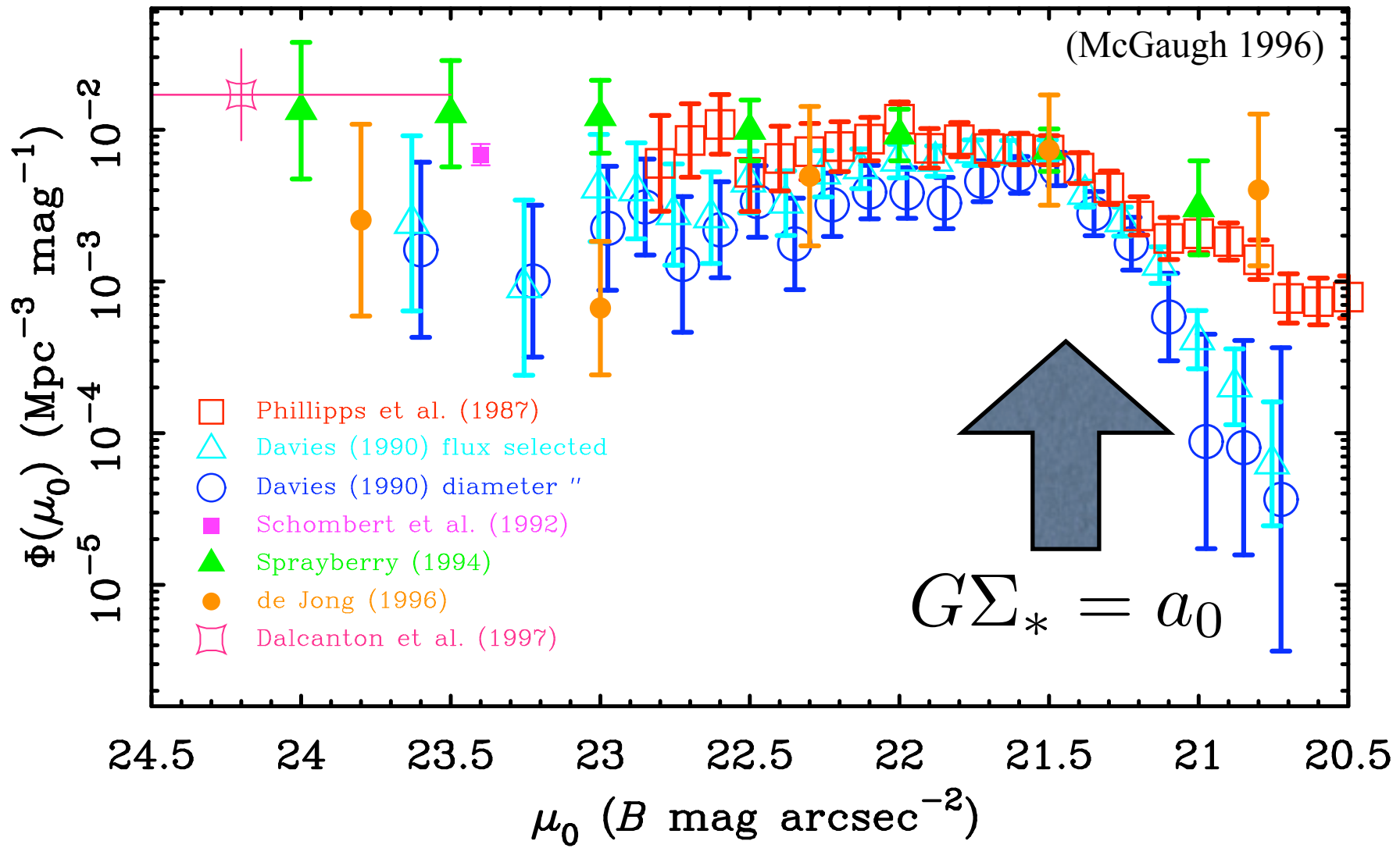

For example, the disk stability limit (Milgrom 1989). This provides a natural explanation for the maximum in the surface brightness distribution (e.g., Freeman's Limit). Disks below the critical surface density are stabilized by MOND; those above are in the Newtonian regime and subject to the usual instabilities. This scale has to be inserted by hand into dark matter models (e.g., Dalcanton et al. 1997). The same scale has been found in SDSS dividing disks and ellipticals (e.g., Kauffmann et al. 2004).

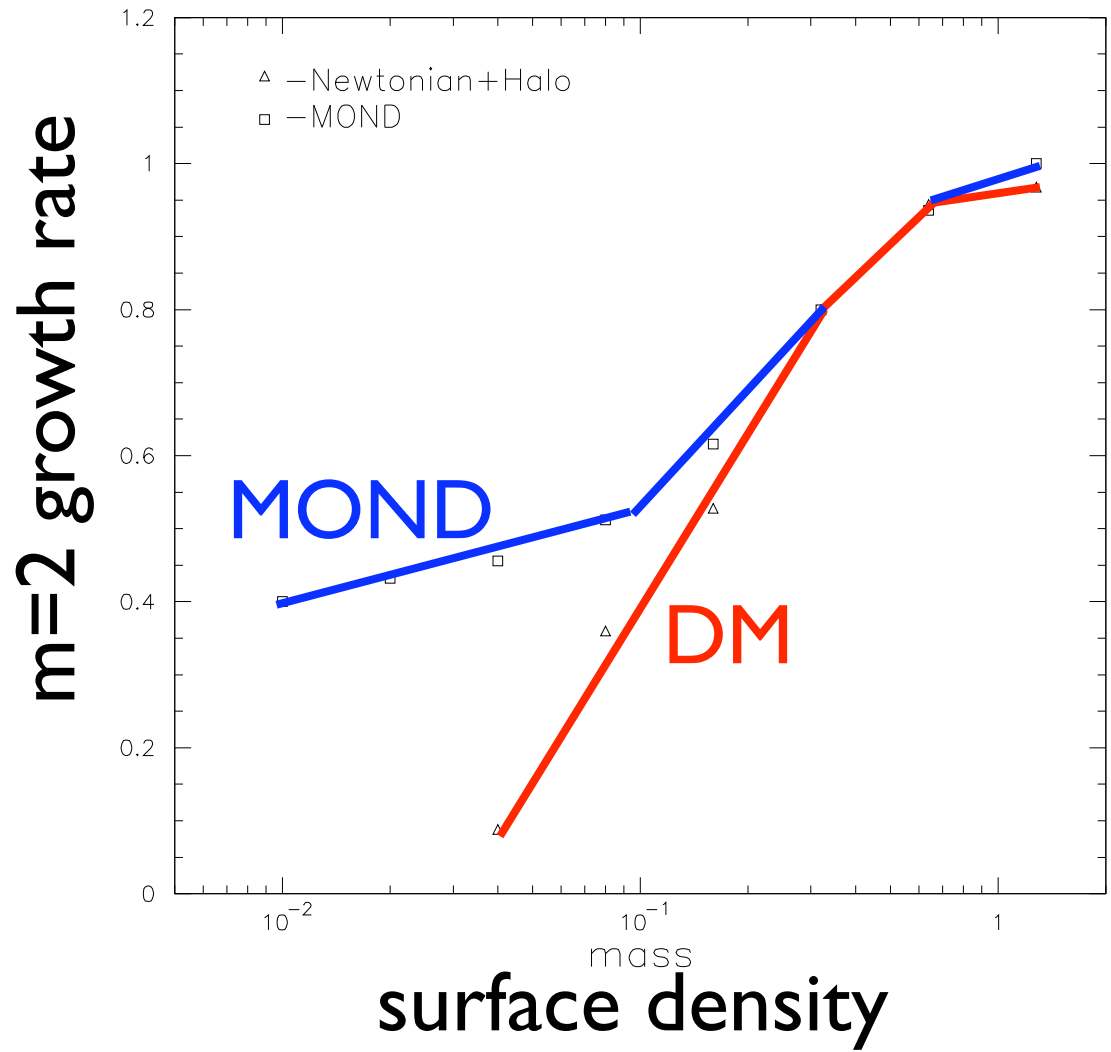

The stability properties of high surface density disks are indistinguishable in dark matter and MOND. However, the two diverge as one goes to low surface density galaxies deep in the MOND regime. The large dark-to-luminous mass ratios in these systems tend to over-stabilize the disks (see also Mihos et al. 1997).

Familiar features like bars and spiral arms are readily understood as disk dynamical features, provided the disk self-gravity is important. The presence of such features in LSB galaxies provides another clue...

Figure 11: The growth rate, in units of the dynamical time, for the m=2 mode as a function of the total mass of the disk. □ MOND, △ Newtonian + Halo. Figure from Brada's Ph.D. thesis (1996). See also Brada & Milgrom (1998).

| m | Q | time step scaling | Growth rate MOND | Newt+DM | halo mass at R=1 |
|---|---|---|---|---|---|

# Disk Masses from Density Waves

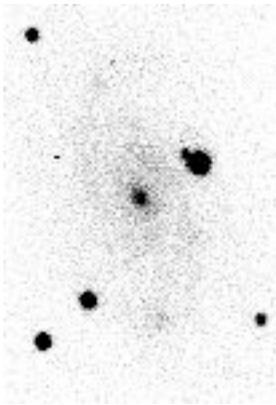

## LSB galaxies got spiral arms!

McGaugh & de Blok (1998) predicted that, if conventional density wave analysis were applied to constrain the mass-to-light ratios of LSB disks, one would infer very high mass-to-light ratios. This follows from the need for disk self gravity to drive spiral features. Subsequent analyses (e.g., Fuchs 2002) found exactly this.

| Galaxy | $(M/L)_*$ |
|---|---|
| F568-1 | 14 |
| F568-3 | 7 |
| F568-6 | 11 |
| F568-V1 | 16 |
| UGC 128 | 4 |
| UGC 1230 | 6 |
| UGC 6614 | 8 |
| ESO 14-40 | 4 |
| ESO 206-140 | 4 |
| ESO 302-120 | 1.7 |
| ESO 425-180 | 2.4 |

Big $(M/L)_*$'s! (F568-1 through UGC 6614)

not LSBs I am familiar with (ESO 14-40 through ESO 425-180)

*from B. Fuchs, astro-ph/0209157*

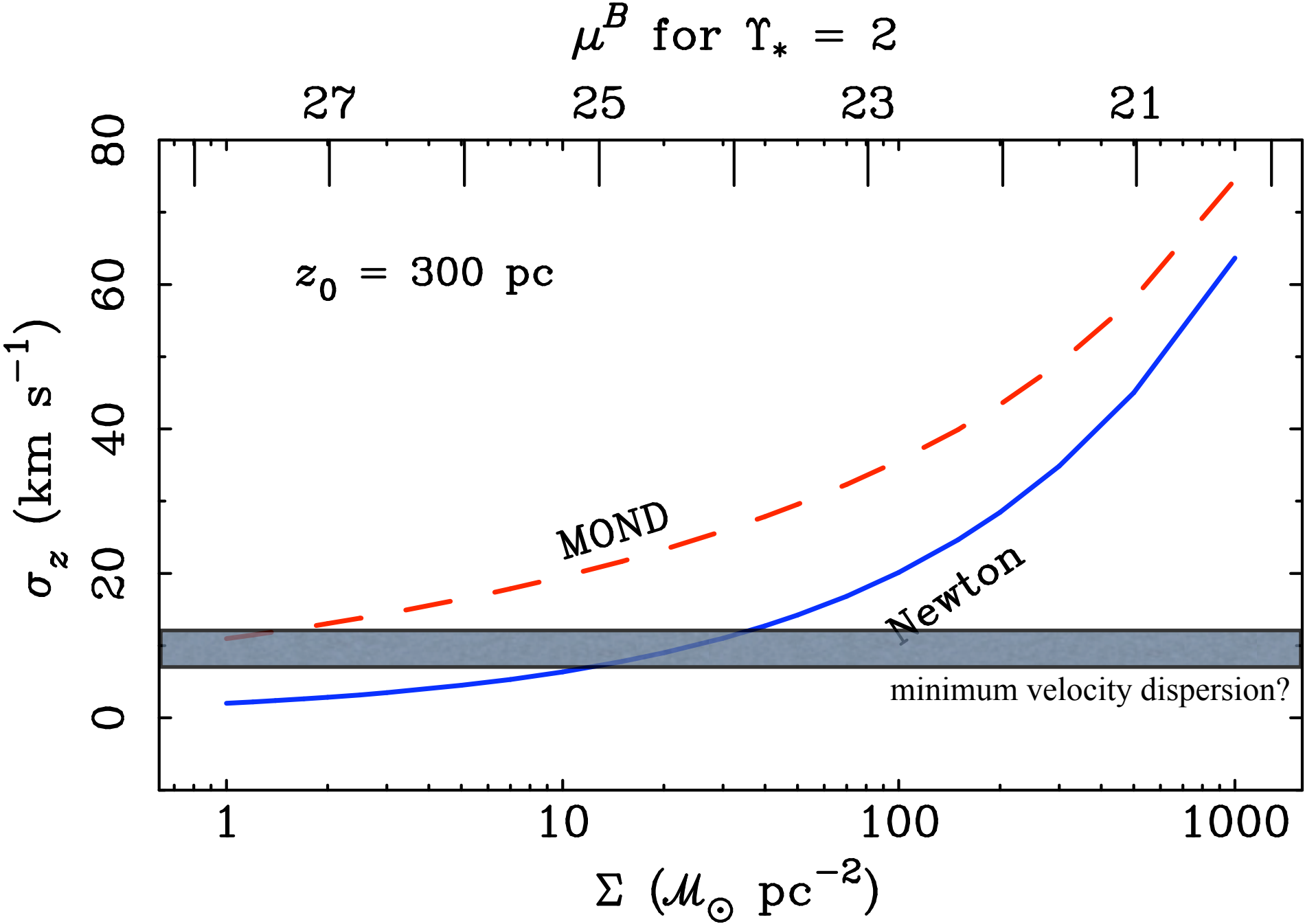

For a disk of a given thickness, MOND can support a higher vertical velocity dispersion than a Newtonian disk plus quasi-spherical dark matter halo. This difference is small at high surface brightness, but becomes pronounced as one goes to low surface densities. MOND provides a natural explanation for the minimum ~7 km/s velocity dispersion frequently measured and for very thin LSB disks seen edge-on.

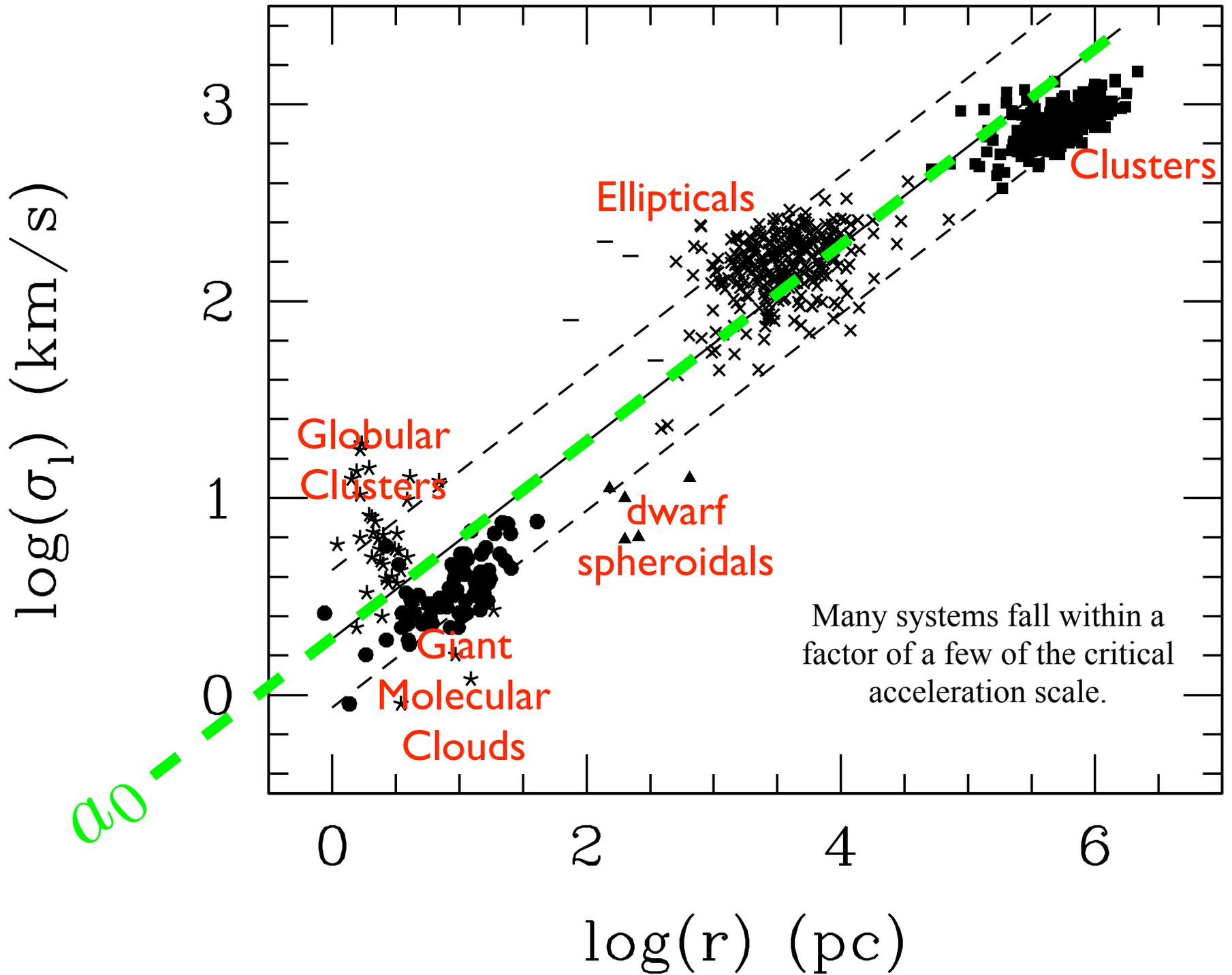

Many systems fall within a factor of a few of the critical acceleration scale.

# dwarf spheroidal satellites of the Milky Way

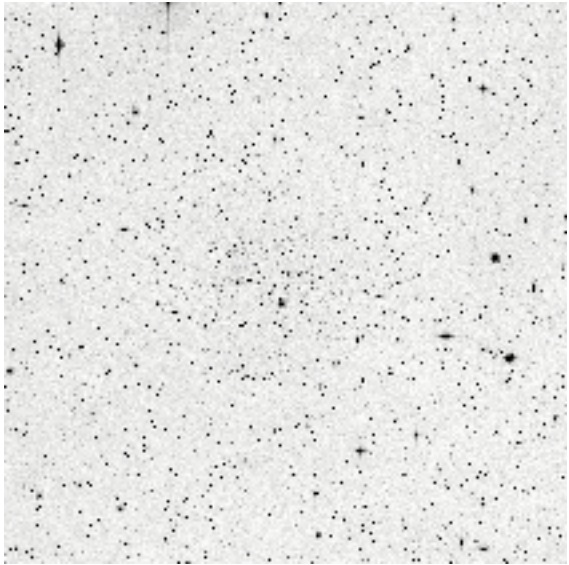

Carina

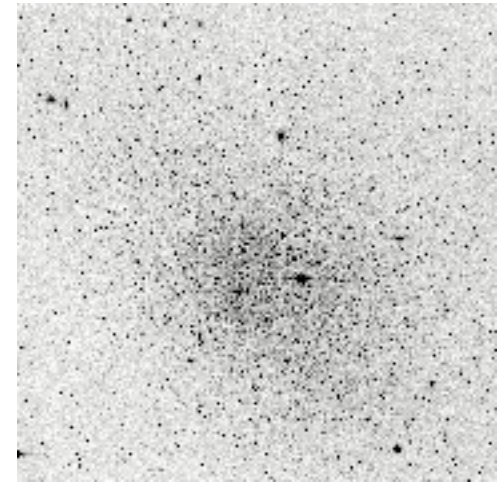

Fornax

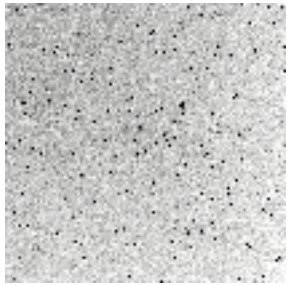

Draco

The dwarf spheroidal satellites of the Milky Way are among the lowest acceleration systems known. As you can see, some of these systems are so low surface density that they are hardly even there!

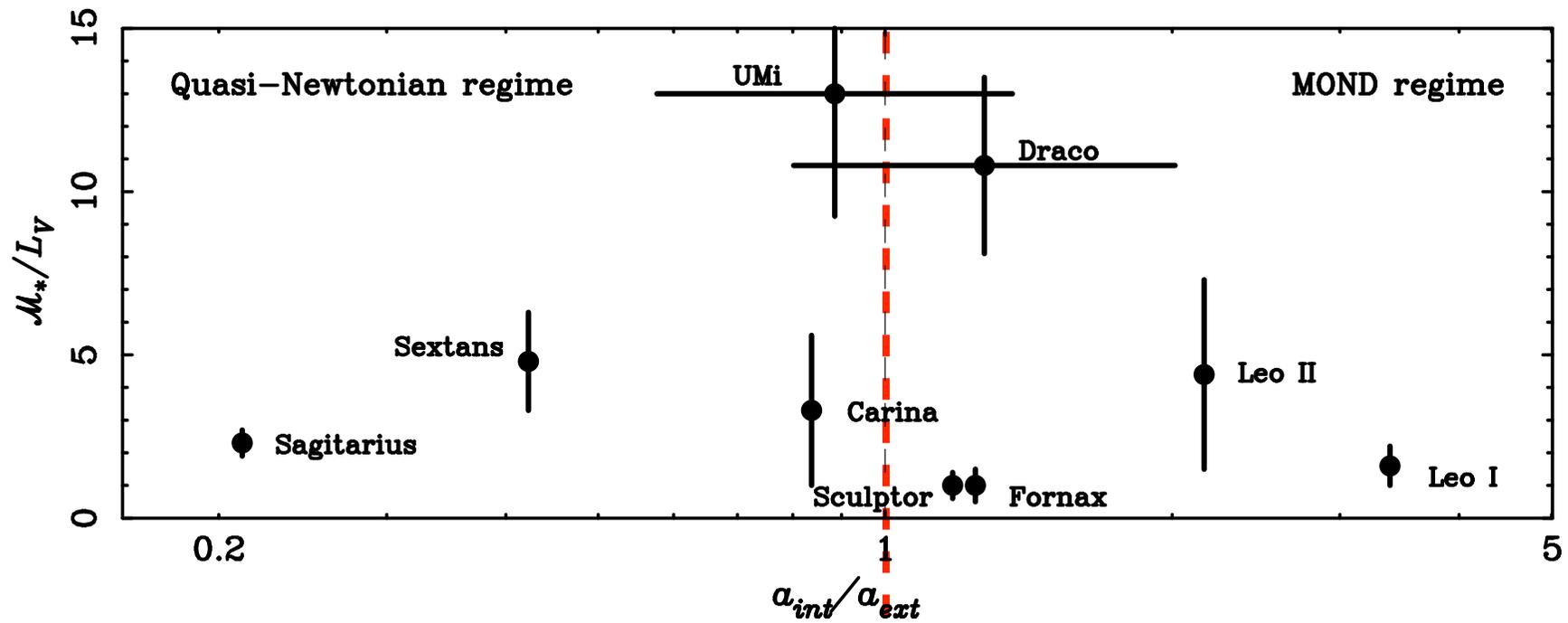

$$\mathcal{M} = \frac{2\sigma^2}{G_{eff} R_V}$$

$$G_{eff} = \frac{G}{\mu(x)}$$

$$\mathcal{M} = \frac{81}{4} \frac{\sigma^4}{a_0 G}$$

The mass estimator in MOND depends on whether the internal field of the dwarf or the external field of the Milky Way dominates. In 7 of 9 cases MOND gives sensible results. In 2 cases it gives M/L too high. In these cases, it is not clear (at 1 sigma) which estimator should be employed. One can spin the interpretation either way here... which is the forest and which are the trees?

# Ellipticals

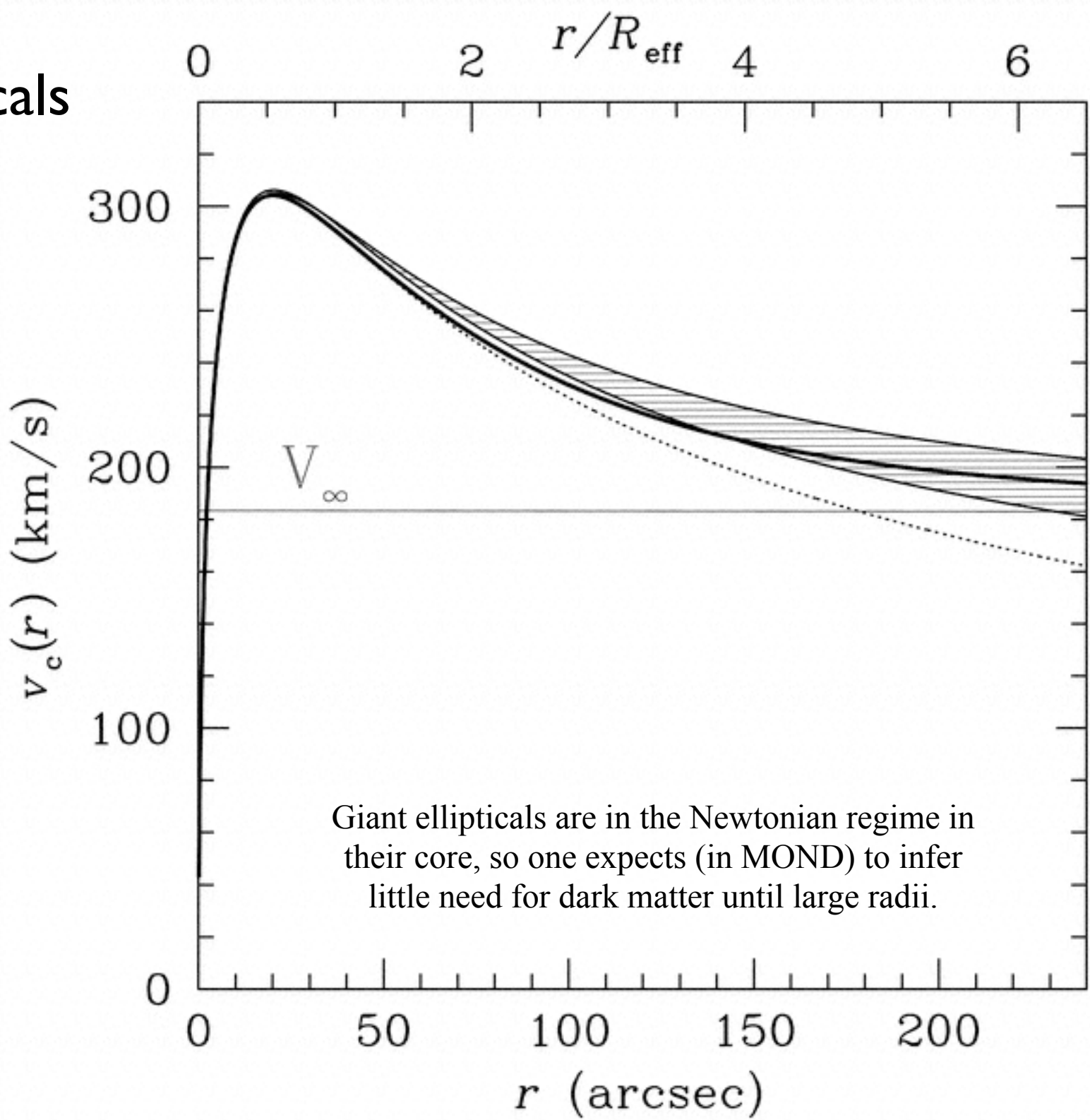

Milgrom (1983); Milgrom & Sanders (2003)

Giant ellipticals are in the Newtonian regime in their core, so one expects (in MOND) to infer little need for dark matter until large radii.

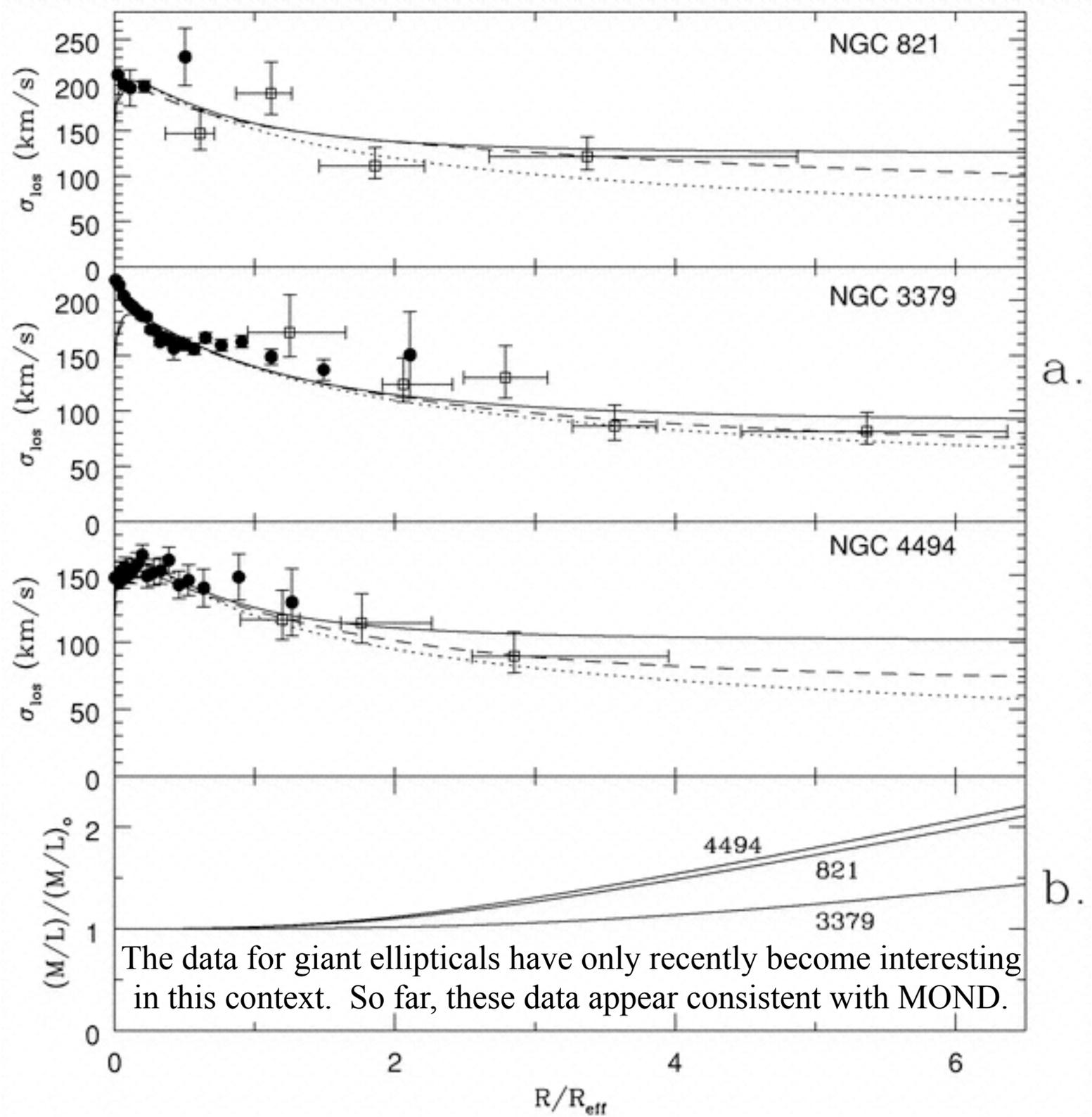

Romanowsky et al. (2003)
Milgrom & Sanders (2003)

The data for giant ellipticals have only recently become interesting in this context. So far, these data appear consistent with MOND.

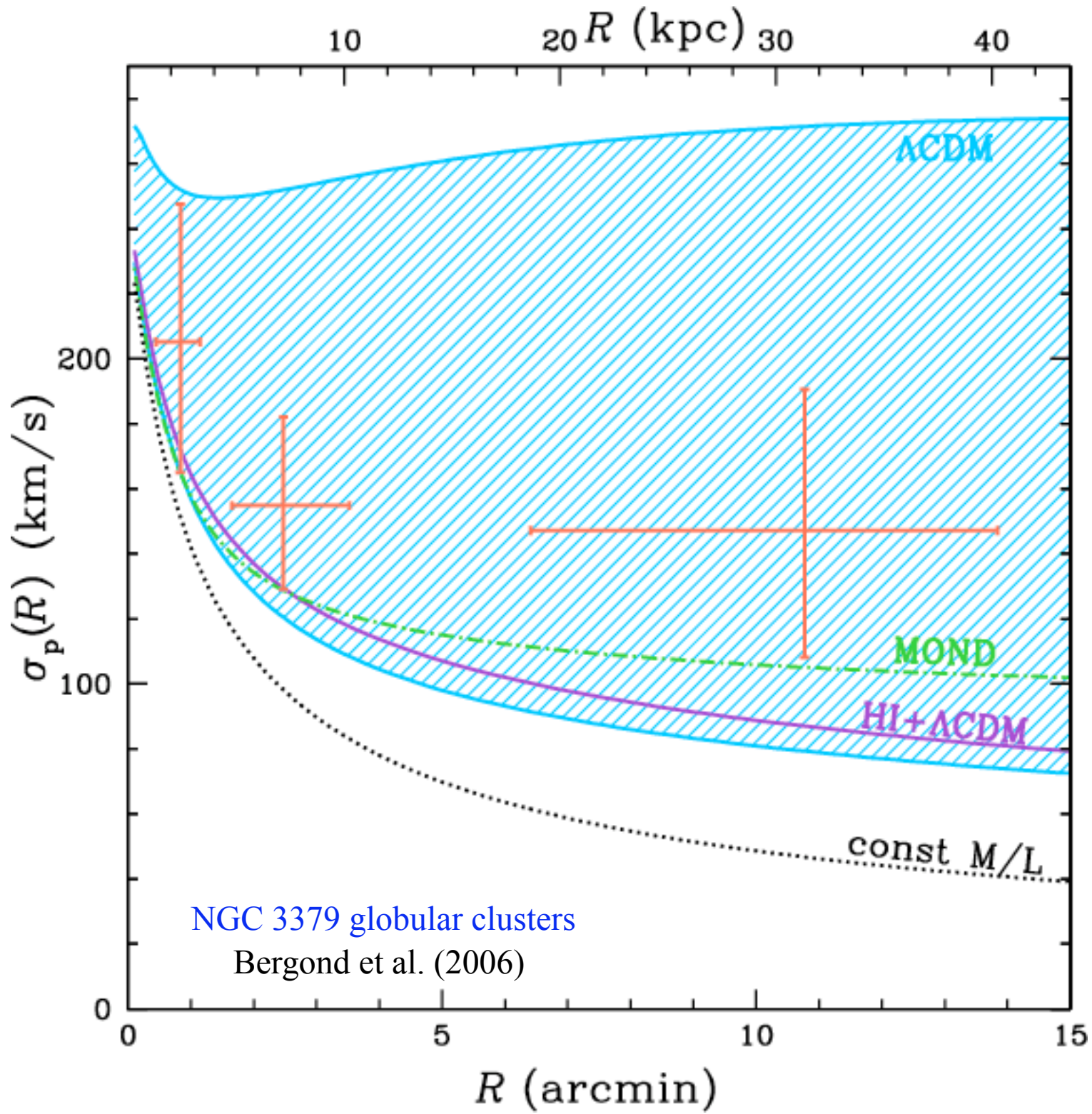

More generally, one wonders why one continues to see the mass discrepancy appear at a particular acceleration scale when ΛCDM predicts such a vast swath of possibilities.

# Clusters of Galaxies

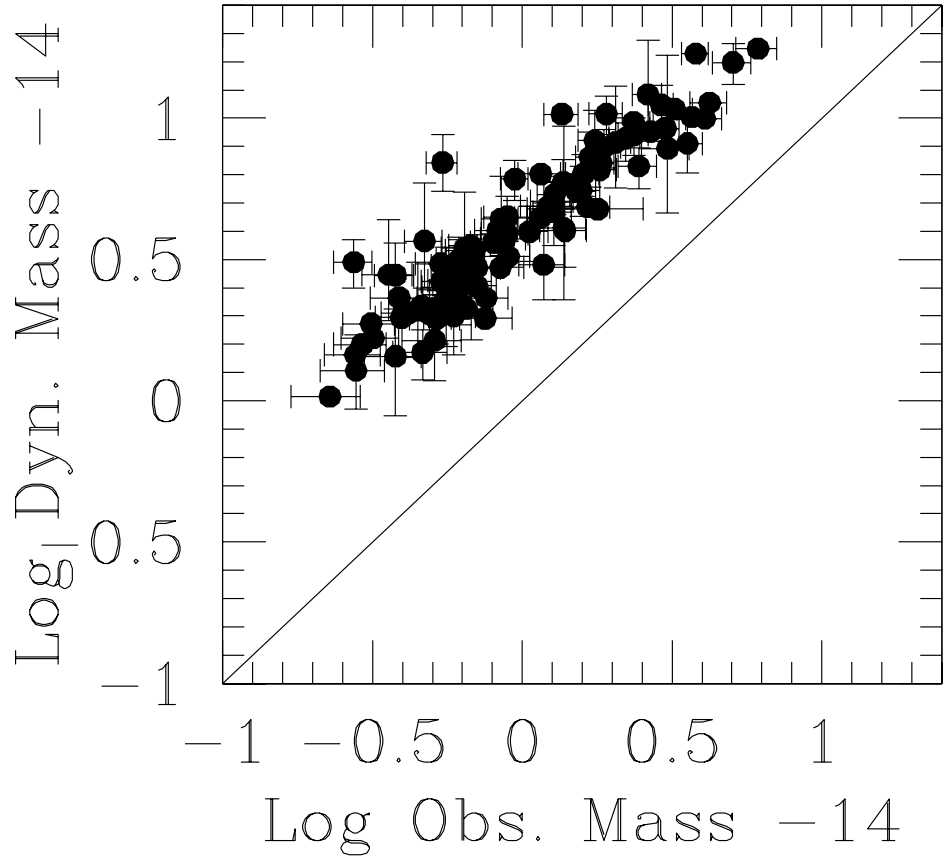
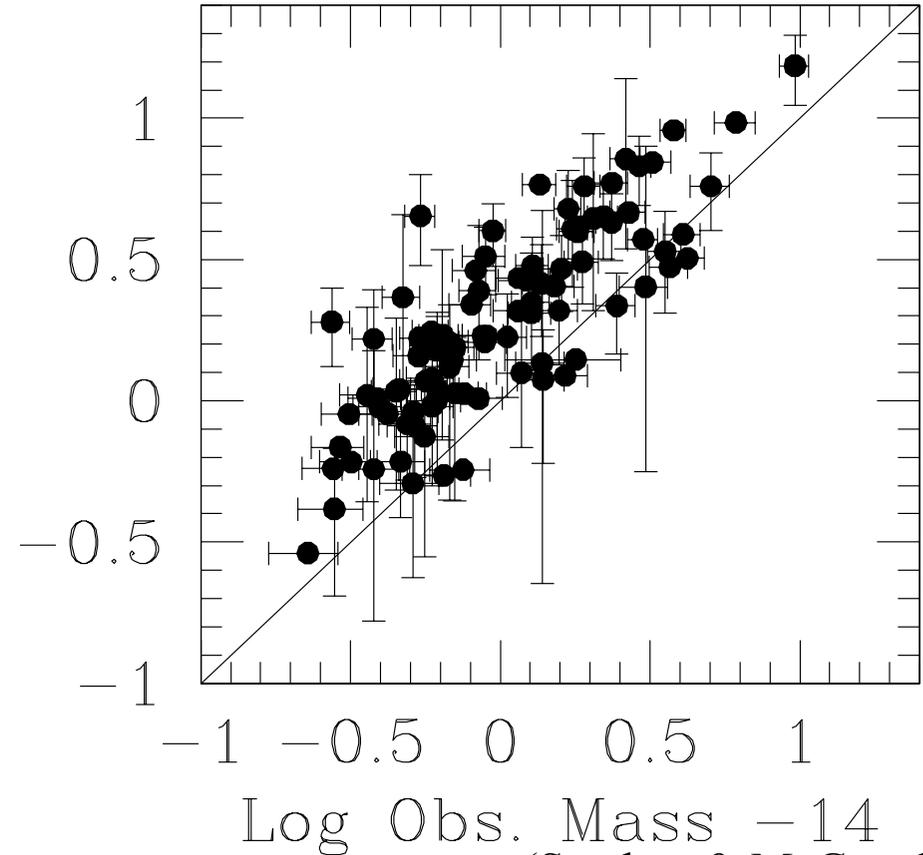

(Sanders & McGaugh 2002)

On the scale of individual galaxies, MOND clearly performs better than CDM.
The opposite is true in rich clusters of galaxies, where MOND does not suffice
to explain the entire mass discrepancy - one needs more mass (neutrinos?).

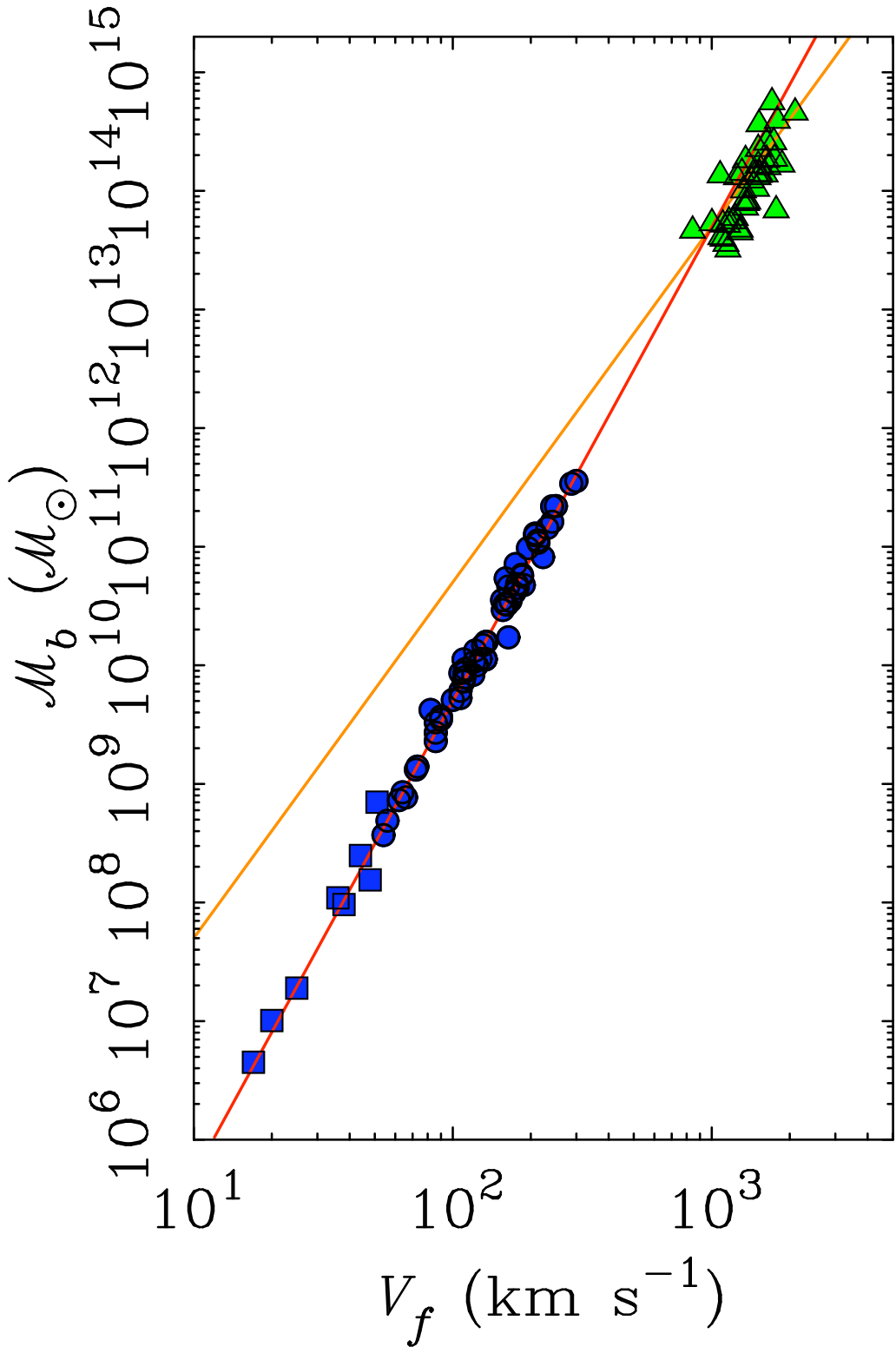

Galaxies and clusters on the same baryonic mass-circular velocity relation. The red line is the prediction of MOND, the orange line that of ΛCDM.

Sanders (2003)
Reiprich (2001)

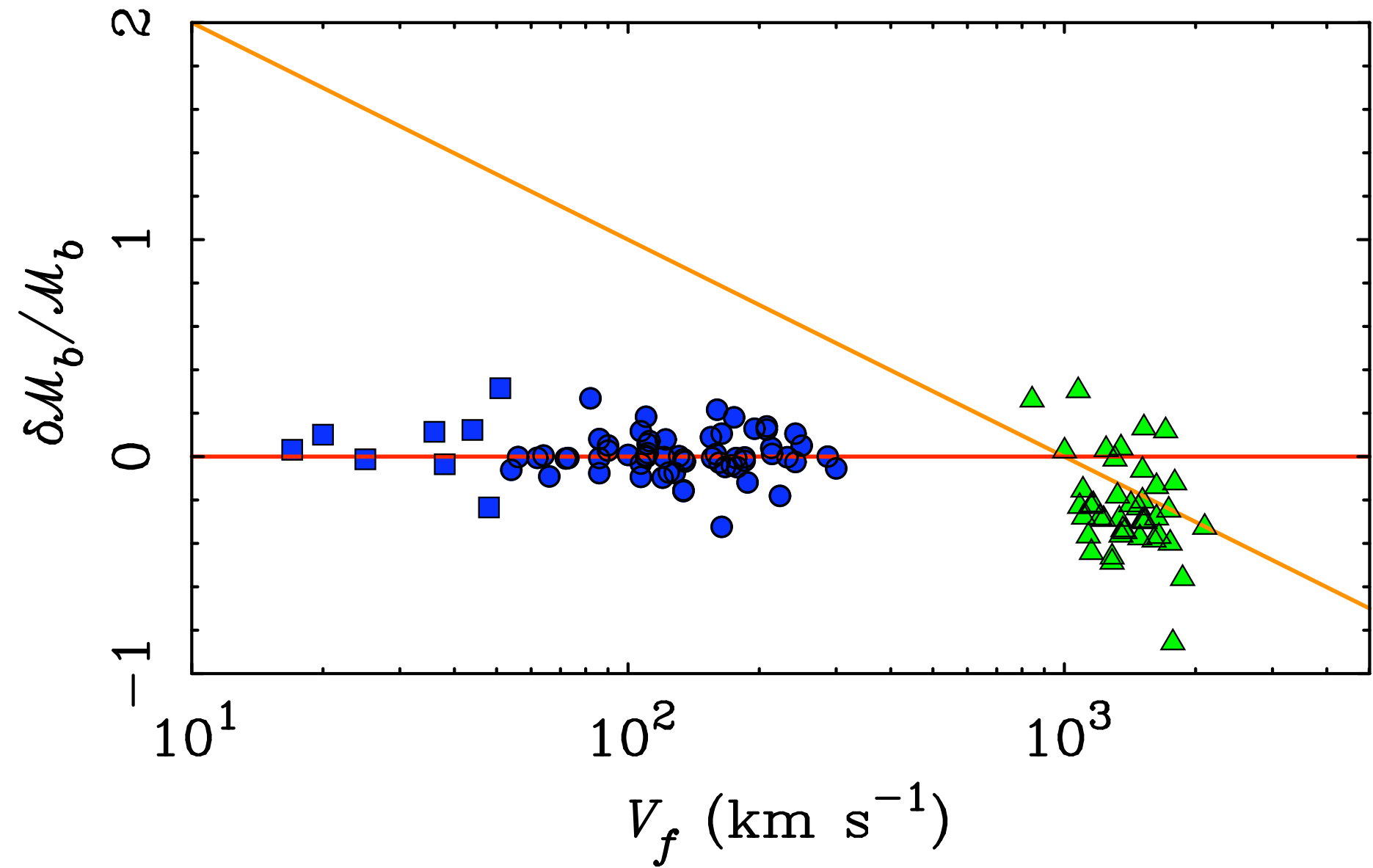

This is the same as the preceding plot but with a slope of 4 taken out. The difference between the red line and the green triangles is the residual mass discrepancy for clusters in MOND. The orange (ΛCDM) line is more consistent with these data. It does rather worse for galaxies. More generally, ~1000 km/s seems to be a break point in the phenomenology. Above this scale, the universe looks like ΛCDM. Below this scale, it looks like MOND.

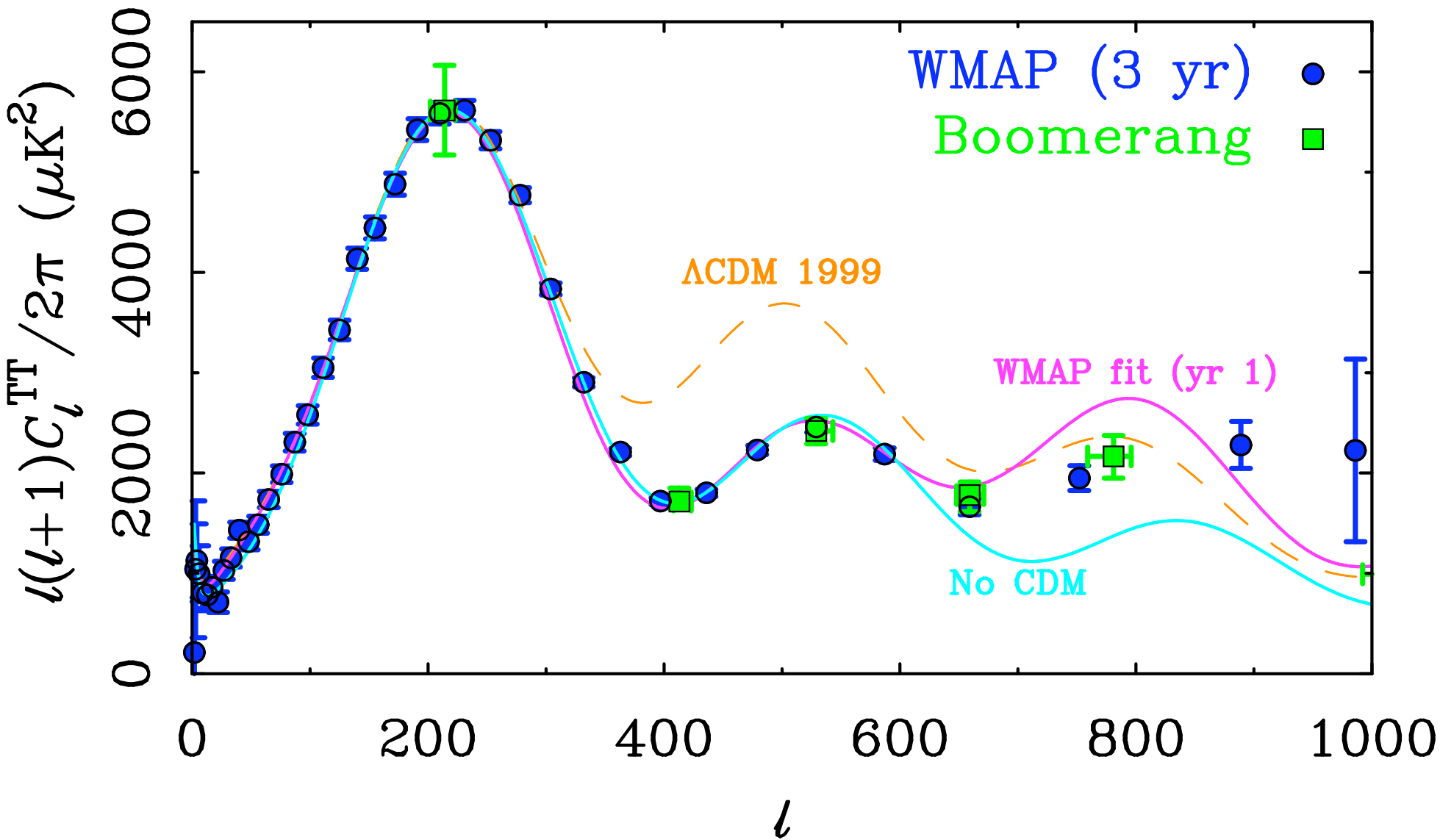

The 3rd year WMAP data compared to various predictions: pre-Boomerang ΛCDM, the fit to the first year WMAP data sans tilt, and no-CDM (McGaugh 1999; 2004). WMAP3 clearly shows power in excess of the low third peak predicted by no-CDM. I thought this would prove fatal to modified gravity theories, but at this conference Skordis & Ferreira showed that I was wrong to think so.

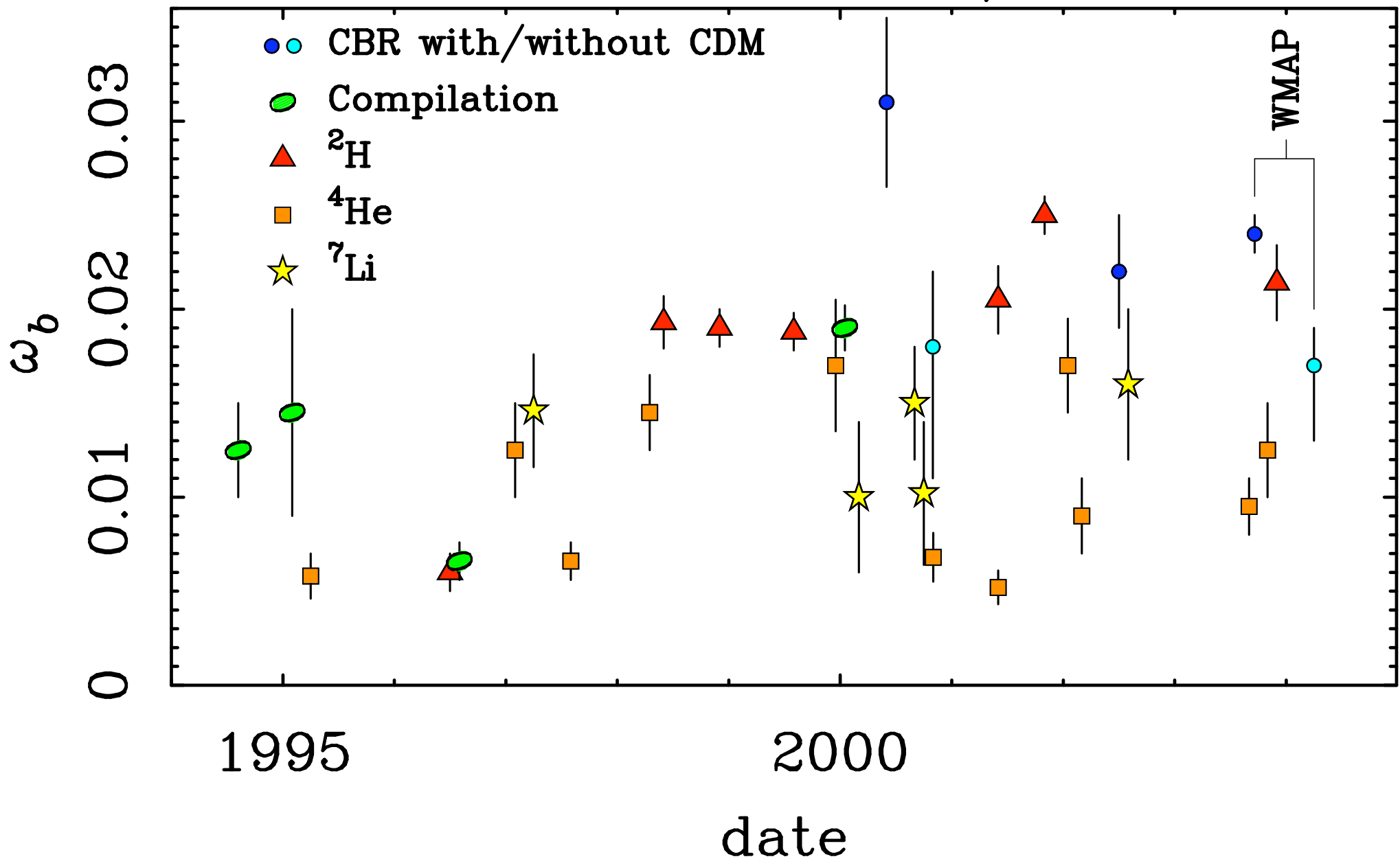

Measurements of BBN abundances (compiled by McGaugh 2004).
The WMAP data without CDM favor a lower baryon density which is more consistent with the bulk of independent data than is the fit with CDM.

Sanders (2001); Sanders & McGaugh (2002)
see also Nusser (2001); Kneib & Gibson (2002)

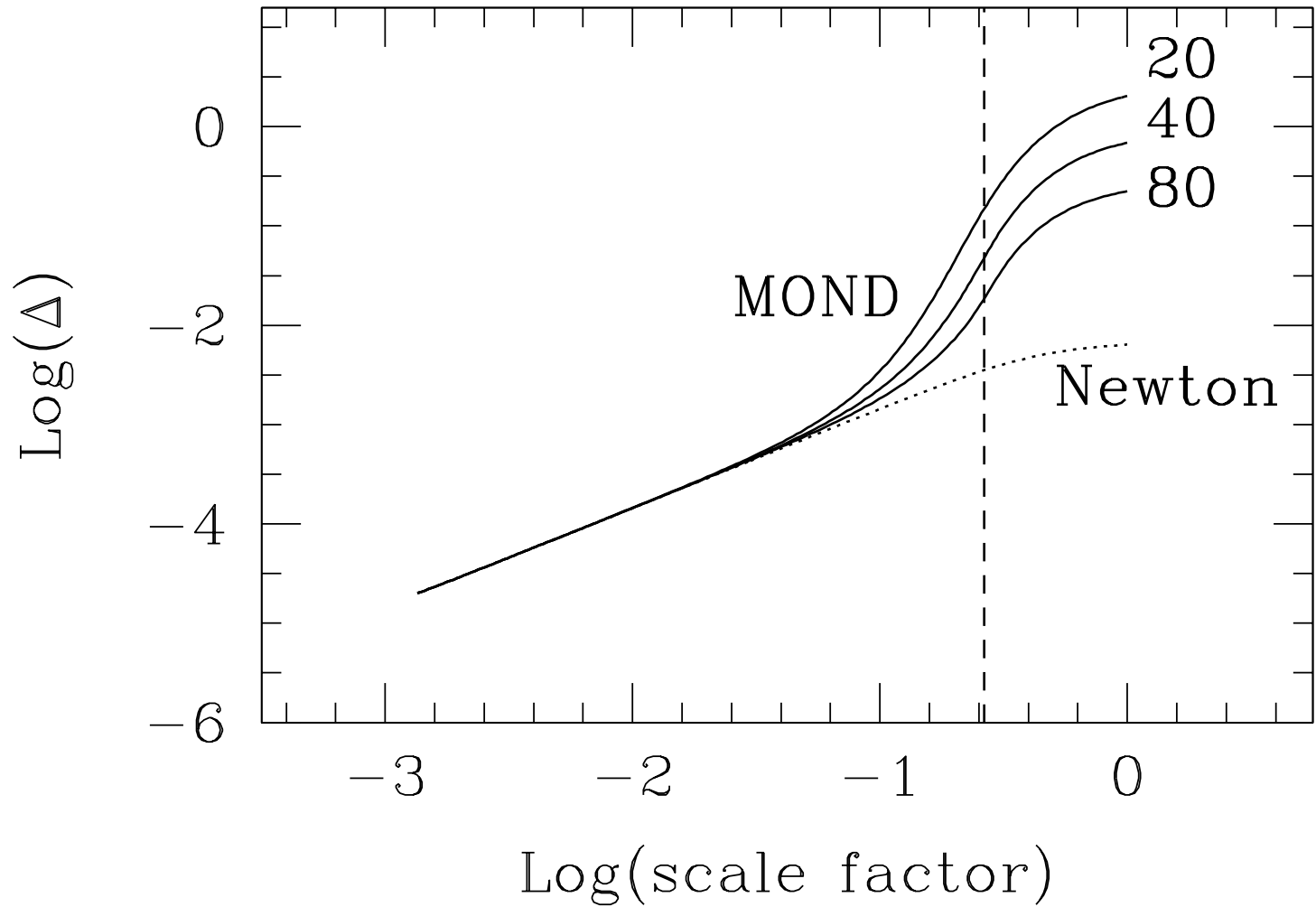

A common misconception is that non-baryonic cold dark matter is required to grow structure. This is only true in the context of GR. If we consider more general theories, the growth rate can be more rapid, potentially achieving the effect usually attributed to dark matter.

# Other MOND tests

- Disk Stability
  - ✔ Freeman limit in surface brightness distribution
  - ✔ thin disks
  - ✔ velocity dispersions
  - ✔ LSB disks not over-stabilized

- ✔ Dwarf Spheroidals ?

- ✔ Giant Ellipticals

- ✘ Clusters of Galaxies

- ? Structure Formation

- Microwave background
  - ✔ 1st:2nd peak amplitude; BBN
  - ✔ early reionization
  - ✔ enhanced ISW effect
  - ✘? 3rd peak ? see Skorids et al. 2005

MOND does well in a variety of tests, not just those with rotation curves. It certainly has problems (e.g., rich clusters; how do galaxies merge?) but it is not obvious that they are worse than those faced by ΛCDM (e.g., not just one but two invisible components; heinous fine-tuning problems to reproduce rotation curve phenomenology). What we need are rigorous predictions that subject theories to falsification. I am most suspicious of theories that claim to fit everything all the time. We should take especial care not to be too credulous of claims that happen to support our most favored theory.